\documentclass[structabstract]{aa} 
\usepackage{natbib}
\usepackage[small,bf,up]{caption} 

\usepackage{graphicx}
\usepackage{amsmath}
\usepackage{txfonts}
\usepackage{hyperref}
\hypersetup{
    pdftitle={S-DIMM+ height characterization of day-time seeing using solar granulation},    
    pdfauthor={G.\ B.\ Scharmer and T.I.M. van Werkhoven},     
    pdfsubject={Solar telescopes, adaptive optics, Site characterization},   
    pdfcreator={pdfLaTeX},   
    pdfkeywords={Solar telescopes, adaptive optics, Site characterization}, 
    pdfnewwindow=true,      
    colorlinks=false,       
    linkcolor=red,          
    citecolor=green,        
    filecolor=magenta,      
    urlcolor=cyan           
}
\newcommand{\refpic}[1]{
Fig.~\ref{#1}%
}
\newcommand{\refeq}[1]{
Eq.~(\ref{#1})%
}

\begin{document}
   \title{S-DIMM+ height characterization of day-time seeing using solar granulation}

\author{G.\ B.\ Scharmer$^{1,2}$ \and T.I.M. van Werkhoven$^{1,3}$}
\authorrunning{Scharmer \& van Werkhoven}
\offprints{\url{scharmer@astro.su.se}}
\institute{Institute for Solar Physics,  Royal Swedish Academy of Sciences, AlbaNova University Center, 
10691 Stockholm \and Stockholm Observatory, Dept. of Astronomy, Stockholm University, AlbaNova 
University Center, 10691 Stockholm \and Sterrekundig Instituut Utrecht, Utrecht University, 
PO Box 80000, 3508TA Utrecht, The Netherlands}

  \abstract
   {To evaluate site quality and to develop multi-conjugative adaptive optics systems for future large solar telescopes, characterization 
   of contributions to seeing from heights up to at least 12~km above the telescope is needed.}
   {We describe a method for evaluating contributions to seeing from different layers along the line-of-sight to the Sun. The method is based on Shack Hartmann wavefront sensor data recorded over a large field-of-view with solar granulation and uses only measurements of differential image displacements from individual exposures, such that the measurements are not degraded by residual
    tip-tilt errors.}
   {The covariance of differential image displacements at variable field angles provides a natural extension of the work of Sarazin and Roddier to include measurements that are also sensitive to the height distribution of seeing. By extending the numerical calculations of Fried to include differential image displacements at distances much smaller and much larger than the subaperture diameter, the wavefront sensor data can be fitted to a well-defined model of seeing. The resulting least-squares fit problem can be solved with conventional methods. The method is tested with simple simulations and applied to wavefront data from the Swedish 1-m Solar Telescope on La Palma, Spain.}
   {We show that good inversions are possible with 9--10~layers, three of which are within the first 1.5~km, and a maximum distance of 16--30~km, but with poor height resolution in the range 10--30~km.} 
   {We conclude that the proposed method allows good measurements when Fried's parameter $r_0$ is larger than about 7.5~cm for the ground layer and that these measurements should provide valuable information for site selection and multi-conjugate development for the future European Solar Telescope. A major limitation is the large field of view presently used for wavefront sensing, leading to uncomfortably large uncertainties in $r_0$ at 30~km distance.} 

   \keywords{Solar telescopes --
                adaptive optics --
                Site characterization
               }

\maketitle

\section{Introduction}
Presently, the most commonly used method for quantifying astronomical seeing is the differential image motion monitor (DIMM), proposed for ESO site testing by \citet{1990A&A...227..294S} and relying on theoretical calculations and suggestions of \citet{1975RaSc...10...71F}. The DIMM built for ESO consists of a single 35~cm telescope with a pupil mask to allow images of bright stars to simultaneously be monitored through two small subapertures separated spatially on a CCD by means of a beam splitter. The advantage of this instrument is that differential image displacements can be measured without impact from telescope tracking errors due to e.g. wind load. By comparing to theoretical calculations of Fried, this allows accurate seeing characterization in terms of the so-called Fried parameter $r_0$ or an equivalent `seeing' disk diameter, related to $r_0$ via $0.98 \lambda/r_0$, where  $\lambda$ is the wavelength. In the following, we refer all values of $r_0$ to a wavelength of 500~nm. 

An instrument similar to the DIMM, the S-DIMM \citep{2001SoPh..198..197L, 2002ASPC..266..350B} was built for site testing for the 4-m Advanced Technology Solar Telescope (ATST), but using one-dimensional differential motion of the solar limb, measured through two 5~cm subapertures. 

A unique advantage of solar observations is the availability of fine structure for wavefront sensing nearly everywhere on the solar surface. This allows large solar telescopes to operate with adaptive optics systems that lock on the same target as observed with science cameras. A particular problem of wavefront sensing with solar telescopes, however, is that low-contrast granulation structures must be usable as targets and this necessitates the use of a fairly large field-of-view (FOV) for wavefront sensing. An important consequence of this is that contributions from higher layers are averaged over an area that increases with height. This degrades the sensitivity to high-altitude seeing.

Within the framework of a design study for the European Solar Telescope (EST), a program for studying two potential sites, one on La Palma and one on Tenerife, has been initiated. The goal is to characterize the height distribution of contributions to seeing at the two sites and to define requirements for a multi-conjugate adaptive optics (MCAO) system for the EST in good seeing conditions. At present, very little work has been published on characterization of high-altitude seeing based on wavefront sensors operating on solar telescopes. Measurements of scintillation of sunlight with a linear array of detectors have been shown to be sensitive to the height distribution of seeing contributions \citep{1993SoPh..145..389S,1993SoPh..145..399B,2002ASPC..266..350B}. Because of the integration of contributions to scintillation over the large solid angle subtended by the solar disk, an array of detectors with fairly large baseline is needed to achieve sensitivity up to a height of only 500 meters \citep{1993SoPh..145..399B}. Instrumentation that utilizes this technique, referred to as `SHAdow BAnd Ranging' \citep[SHABAR;][]{2001ExA....12....1B} was built to characterize near ground seeing in connection with site testing for the ATST \citep{2006SPIE.6267E..59H}. A similar instrument, but with much longer baseline (3.2~m) than that used for the ATST, is under construction for EST site testing (Collados, private communication). 

The first published Shack-Hartmann (SH) based measurements of high-altitude seeing with solar telescopes are those of Waldmann et al. (\citeyear{2008SPIE.7015E.154W}; 2010 in preparation), based on the covariance of local image displacements from different subapertures of a wide-field wavefront sensor (WFWFS). However, methods that are based on covariances (or correlations) of {\em absolute} image displacements are sensitive to tracking errors and vibrations in the telescope \citep{1975RaSc...10...71F}. Waldmann et al. (\citeyear{2008SPIE.7015E.154W}; 2010, in preparation) therefore also implemented modeling of compensation of such image displacements with a tip-tilt mirror and adaptive optics (AO), assumed to correct Zernike aberrations. A technique related to that proposed below is the SLODAR \citep{2002MNRAS.337..103W} for night-time measurements of seeing. The principle of this instrument is to use the cross-correlation of image motion measured for binary stars with a SH wavefront sensor through different subapertures. To eliminate the impact of telescope guiding errors, the average image motion of both stars measured with all subapertures is subtracted from the image motions of individual subapertures for each exposure separately. However, this procedure also removes the common atmospheric wavefront tilt, averaged over the entire telescope aperture for each of the two stars observed. This introduces an anisoplanatic component of the wavefront \citep{2006MNRAS.369..835B}. As shown by the same authors, this can be compensated for in the analysis of the data, but at the price of making the analysis considerably more complex than in the originally proposed method by \citet{2002MNRAS.337..103W}.

In the present paper, we propose a method that is also insensitive to image motion from the telescope or residual errors from a tip-tilt mirror but quite simple to model. The present method relies on measurements of the covariance of {\em differential} image displacements (between two subapertures) at different field angles. As we shall see, this approach can be considered as a natural extension of the DIMM and S-DIMM approach to include measurements that are sensitive to the height variation of seeing and we therefore refer to the proposed instrument as S-DIMM+. We show that S-DIMM+ data can be inverted using calculations made by \citet{1975RaSc...10...71F} to model contributions from different layers along the line of sight (LOS). The paper is organized as follows: In Sect. 2, we describe the assumptions made, the method and extend the calculations made by \citet{1975RaSc...10...71F}. In Sect. 3, we describe the optical setup and the algorithms used to measure differential image displacements and to compensate for noise bias. In Sect. 4, we describe the results of the first data obtained with the S-DIMM+ installed at the Swedish 1-m Solar Telescope (SST) on La Palma and in Sect. 5 we give concluding remarks about the proposed method and its limitations.

\section{Method}
We make the following assumptions:
\begin{enumerate}
\item A measurement of image position is equivalent to a measurement of the slope of the wavefront over the subaperture.
\item A measurement of image position averaged over the FOV used is equivalent to averaging the wavefront Zernike tip or tilt over the solid angle corresponding to the subaperture and the FOV.
\item The atmosphere is modeled as consisting of $N$ discrete (thin) layers.
\item The contributions from different layers are statistically independent of each other.
\item The turbulent fluctuations in refractive index giving rise to the inferred wavefront slopes are statistically homogeneous and isotropic within a given layer. These fluctuations can be characterized in terms of a structure function based on Kolmogorov turbulence.
\item Propagation and saturation effects are negligible. For a discussion of these effects on DIMM measurements, see \citet{2007MNRAS.381.1179T}.
\end{enumerate}

\begin{figure}%
   \centering
   \includegraphics[bb=2 3 215 210, clip, width=0.4\textwidth]{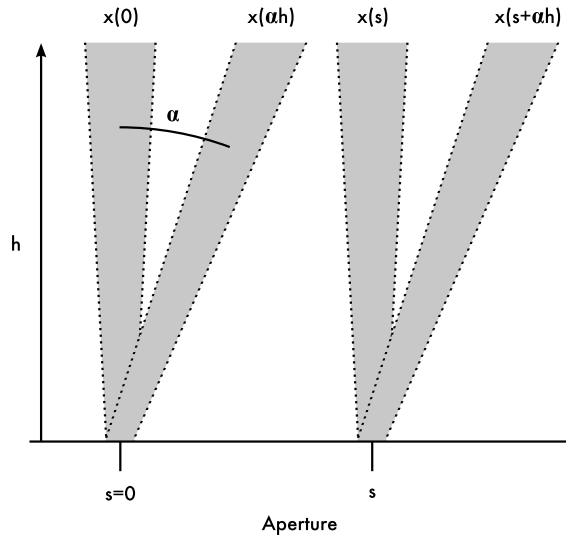}
   \caption{\label{fig:1}Layout of the relation between the wavefront sensor geometry and the contributions to the two differential wavefront slope (image displacement) measurements from a height $h$. The separation between the two subapertures (indicated as heavy black lines) is $s$ and the relative field angle between the two sub-fields measured is $\alpha$. Note the divergence of the `beams' with height due to the non-zero FOV used for these wavefront measurements.}
\end{figure}

The present method relies on using {\em individual} exposures to measure {\em relative} image displacements of the \textit{same} target observed through different subapertures. These restrictions are essential:By measuring differential image displacements from an individual exposed image, errors from telescope movements are eliminated; by using the {\em same target} observed through different subapertures we can use cross-correlation techniques with arbitrary solar fine structure to measure relative image displacements. We shall use the {\em covariance} of two such 
\textit{differential} measurements, taken at different field angles, to achieve our goal of characterizing seeing. Furthermore, we shall restrict the present analysis to measuring only relative image displacements that are either along (longitudinal or x-component) or perpendicular (transverse or y-component) to the line connecting the centers of the two subapertures. Because of the assumed isotropy and homogeneity of the turbulent seeing fluctuations, the problem is then one-dimensional such that the measured relative image displacements can only depend on the distance between the two measurement points. Without loss of generality, we can assume that one of the subapertures is located at the origin, whereas the second subaperture is located at a distance $s$ from that subaperture. Furthermore, we assume that the field angle of the first measurement is zero, while that of the second measurement is $\alpha$ and that this tilt is along the x-component of the image displacements. For the first measurement, the measured quantity
$\delta x_1$ then corresponds to added contributions from the $N$ layers, located at heights $h_n$\footnote{We will consistently refer to the heights $h_n$ of seeing layers above the telescope as if observations were made with the Sun at zenith. For observations made at a zenith distance $z$, heights appearing in equations below need to be divided by $\cos z$ to correspond to the actual distance between the telescope and the seeing layer.}: 
\begin{equation}%
\label{eq:1}\delta x_1(s, 0) = \sum\limits_{n=1}^{N} (x_n(s) - x_n(0)) , 
\end{equation}%
and that of the second measurement corresponds to
$\delta x_2$ is then  
\begin{equation}%
\label{eq:2}\delta x_2(s, \alpha) = \sum\limits_{n=1}^{N} (x_n(s+\alpha h_n) - x_n(\alpha h_n)) .
\end{equation}%
Because of the assumed independence of contributions from different layers, the covariance between $\delta x_1$ and $\delta x_2$ is given by%
\begin{equation}%
\label{eq:3}\langle\delta x_1 \delta x_2\rangle  = \sum\limits_{n=1}^{N} \langle (x_n(s) - x_n(0))~(x_n(s+\alpha h_n) - x_n(\alpha h_n)) \rangle ,
\end{equation}%
where $\langle ... \rangle$ denotes averages over many exposed frames.
Evaluating the four terms one by one and taking advantage of the assumed homogeneity of the statistical averages at each height $h_n$, we can rewrite \refeq{eq:3} as a combination of three variances of \textit{differential} image displacements
\begin{multline}%
\label{eq:4}\langle  \delta x_1 \delta x_2)\rangle  = \sum\limits_{n=1}^{N} \langle (x_n(\alpha h_n -s)-x_n(0))^2\rangle /2 \\
    + \langle (x_n(\alpha h_n+s)-x_n(0))^2\rangle /2 - \langle (x_n(\alpha h_n)-x_n(0))^2\rangle .
\end{multline}%
This form of the equation clearly shows the connection to DIMM \citep{1990A&A...227..294S} and S-DIMM measurements. When $\alpha=0$, the last term disappears, the first two terms are equal and we have a conventional DIMM, except that the contributions from the high layers are reduced by the averaging of image displacements from the relatively large FOV (see below).

\citet{1990A&A...227..294S} gave an approximate  equation for estimating the variance of differential image displacements recorded with two subapertures of diameter $D$, separated by a distance $s$:
\begin{equation}%
\label{eq:5}\langle (x(s)-x(0))^2\rangle  = 0.358 \lambda^2 r_0^{-5/3} D^{-1/3} (1 - 0.541 (s/D)^{-1/3}) ,
\end{equation}%
for longitudinal image displacements and 
\begin{equation}%
\label{eq:6}\langle (y(s)-y(0))^2\rangle  = 0.358 \lambda^2 r_0^{-5/3} D^{-1/3} (1 - 0.811 (s/D)^{-1/3}) ,
\end{equation}%
for transverse image displacements. Here, $D$ is the subaperture diameter and $r_0$ is Fried's parameter. Adopting their notation and that of \citet{1975RaSc...10...71F}, we rewrite these equations as
\begin{eqnarray}%
\label{eq:7}\langle (x(s)-x(0))^2\rangle  = 0.358 \lambda^2 r_0^{-5/3} D^{-1/3} I(s/D,0), \\
\label{eq:8}\langle (y(s)-y(0))^2\rangle  = 0.358 \lambda^2 r_0^{-5/3} D^{-1/3} I(s/D,\pi /2) ,
\end{eqnarray}%
where it is assumed that the two subapertures are separated along the x-axis.
The function $I$ is normalized such that it approaches unity when $s$ approaches infinity and is symmetric $I(-s/D,0)=I(s/D,0)$ and $I(-s/D,\pi /2)=I(s/D,\pi /2)$. With an assumed zero inner scale and an infinite outer scale for the turbulence, the function $I$ can only depend on separation $s$, divided by the diameter $D$ of the averaging (round) aperture. The approximations given by Sarazin and Roddier are reasonably accurate only for $s>D/2$. However, to implement the present method we need more accurate estimates of $I$ also when $s/D$ approaches zero and we need to take into account the averaging effect of using a large FOV for wavefront sensing (see below). We have extended the calculations of \citet[his Eq. (32)]{1975RaSc...10...71F} of $I(s/D,0)$ and $I(s/D,\pi /2)$ to include also $s/D < 1$ and $s/D$ as large as 50, corresponding to a distance of 5~m with 10~cm subapertures. The results of these calculations agree perfectly with those of Fried when $s/D=1$ but systematically deviate with increasing $s/D$ and differ by 10\% when $s/D=10$. As suspected by \citet{1990A&A...227..294S}, this suggests numerical inaccuracies in the calculation of Fried. We confirm this by noting that the calculations of Fried show a slight {\em decrease} of longitudinal differential image displacements when $s/D > 7.5$, which is clearly not physical. We obtain similar behavior in our calculations when the step size of one of the integrating variables ($u$ in Fried's Eq. 32) is too large and have decreased that step size by approximately a factor of 80 compared to calculations essentially reproducing the results of Fried. With this improvement, the calculations show monotonous increase of $I(s/D,0)$ and $I(s/D,\pi /2)$ with $s$ as long as $s/D < 300$ which is a much larger range than needed for the present analysis. We therefore believe that our calculations must be more accurate than those of \citet{1975RaSc...10...71F}. 

For wavefront sensing with an extended target, the averaging area corresponds to the sub-pupil area only close to the telescope. \textit{The effective averaging area expands from the pupil and up by an amount that increases with the FOV} used for wavefront sensing. A reasonable estimate of the effective diameter $D_{\rm eff}$ can be obtained by calculating the convolution of the binary aperture and the FOV used for wavefront sensing projected at the height $h$. At small heights, the convolved function will be unity over an area corresponding to the sub-pupil and gradually fall off outside the sub-pupil. By calculating the area of the convolved binary pupil and FOV and equating that to $\pi D_{\rm eff}^2/4$, an {\em effective} diameter $D_{\rm eff}$ can be defined. A good approximation is to set $D_{\rm eff}$ to the maximum of $D$ and $\phi h$, where $\phi$ is the (average) diameter of the FOV. At large heights, $D_{\rm eff}$ is much larger than the sub-pupil diameter $D$. At a distance of 30~km and with a FOV diameter of $\phi=5.5$~arcsec, we have $\phi h_n=0.8$~m, which is 8 times larger than the subaperture diameter $D=0.098$~m used for recording data. The averaging effect of the large FOV used thus has strong effect on the determination of $r_0$ from differential image displacements measurements for the higher layers. In particular, we should not expect the method to work well, or even at all, when $D_{\rm eff}$ is larger than $r_0$. This `cone effect' is similar to that of measurements of the integrated scintillation from the entire solar disk \citep{1993SoPh..145..399B} but of much reduced magnitude since the SH measurements use a typical FOV of 5--6~arcsec diameter, whereas the solar disk subtends a diameter 30 times larger.

\begin{figure}[t]%
   \centering
   \includegraphics[width=0.45\hsize]{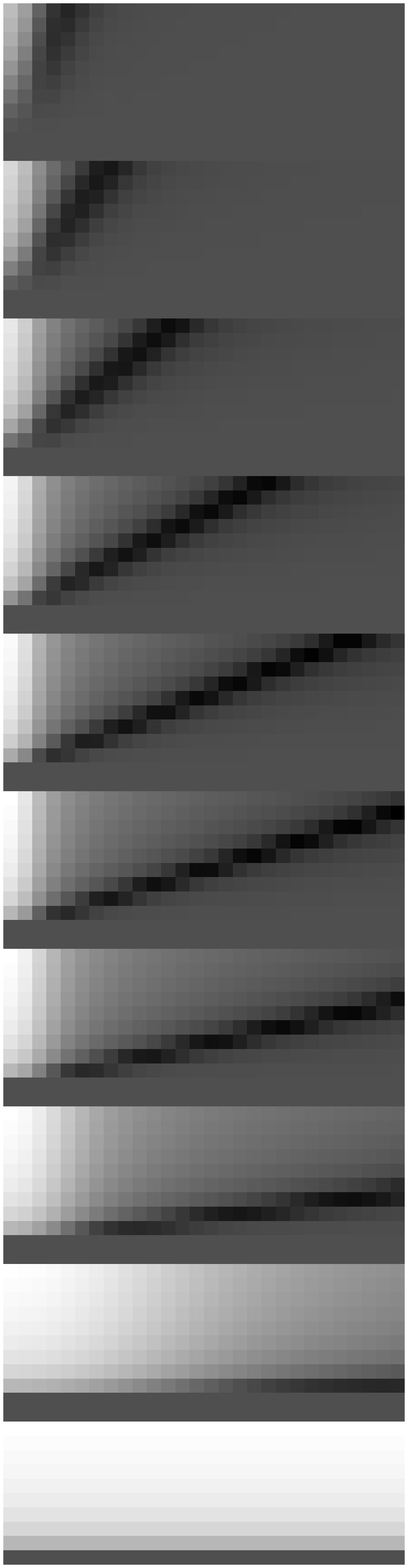}
   \includegraphics[width=0.45\hsize]{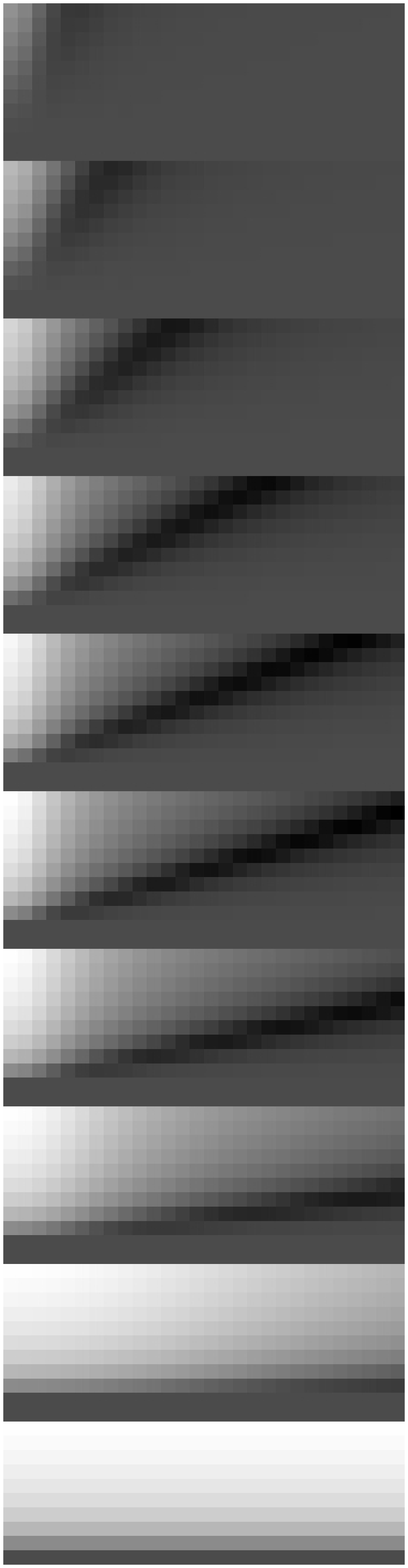}
   \includegraphics[width=0.50\hsize]{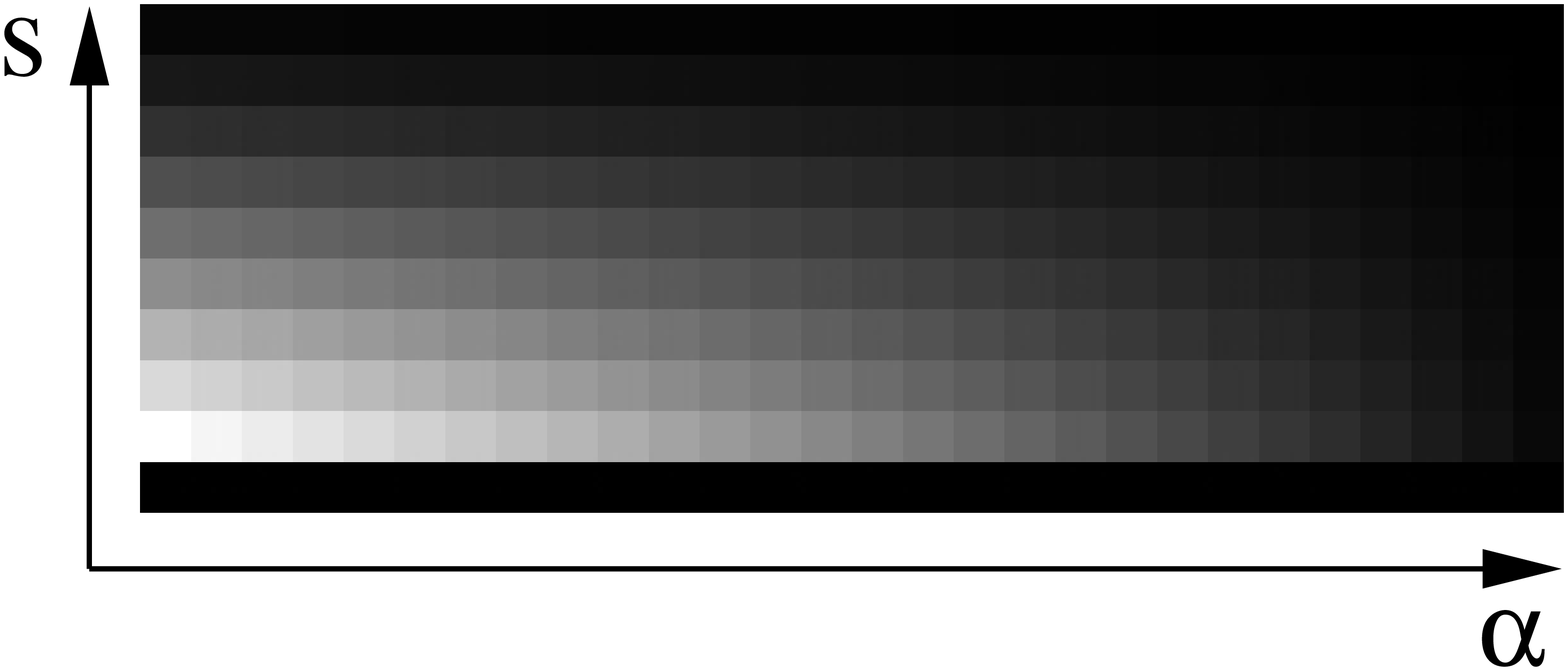}
   \caption{\label{fig:2}Theoretical covariance functions, given by Eqs. (\ref{eq:11}) and (\ref{eq:12}), as function of separation $s$ between the subapertures (increasing upwards in each sub-panel) and field angle difference $\alpha$ (increasing from left to right). The left set of panels shows the covariance for longitudinal differential image displacements, the right panel that of transverse differential image displacements. The covariance functions have been calculated for heights 0.0, 0.5, 1.5, 2.5, 3.5, 4.5, 6.0, 9.5, 16 and 30~km and are displayed in order from bottom to top. The lowermost panel shows the number of measurements for each $(s,\alpha)$, applied as a weight $W(s,\alpha)$ in the least-squares fitting procedure. The lowermost row in all panels corresponds to $s=0$, for which the covariance functions are zero.}
\end{figure}

We conclude that we can use the theory developed by \citet{1975RaSc...10...71F} for modeling differential image displacements measured with the proposed method by simply replacing $s$ with $s+\alpha h$ and the pupil diameter $D$ with an effective diameter $D_{\rm eff}$. To make the averaging area roundish, we simply apply an approximately round binary mask when measuring image displacements from the granulation images with cross-correlation techniques. Combining Eqs. (\ref{eq:4}), (\ref{eq:7}) and (\ref{eq:8}), we now obtain%
\begin{equation}%
\label{eq:9}\langle  \delta x_1 \delta x_2)\rangle = \sum\limits_{n=1}^{N} c_n F_x(s,\alpha, h_n) ,
\end{equation}%
\begin{equation}%
\label{eq:10}\langle  \delta y_1 \delta y_2)\rangle  = \sum\limits_{n=1}^{N} c_n F_y(s,\alpha, h_n) ,
\end{equation}%
where%
\begin{multline}%
\label{eq:11}F_x(s,\alpha, h_n)=I((\alpha h_n-s)/D_{\rm eff},0)/2 \\
+ I((\alpha h_n+s)/D_{\rm eff},0)/2 - I(\alpha h_n/D_{\rm eff},0) ,
\end{multline}%
\begin{multline}%
\label{eq:12}F_y(s,\alpha, h_n)=I((\alpha h_n-s)/D_{\rm eff},\pi /2)/2 \\
+ I((\alpha h_n+s)/D_{\rm eff},\pi /2)/2 -I(\alpha h_n/D_{\rm eff},\pi/2) .
\end{multline}%
The coefficients $c_n$ are given by%
\begin{equation}%
\label{eq:13}c_n= 0.358 \lambda^2 r_0(h_n)^{-5/3} D_{\rm eff}(h_n)^{-1/3} ,
\end{equation}%
expressed in terms of Fried's parameter $r_0$. We can also express the results in terms of the turbulent strength of each layer, $C_n^2 dh$, %
\begin{equation}%
\label{eq:14} C_n^2 dh = 0.0599 \cos(z) \lambda^2 r_0^{-5/3},
\end{equation}%
where $C_n^2(h_n)$ is the atmospheric structure constant and $dh$ is the thickness of each (thin) layer. We obtain the relation%
\begin{equation}%
\label{eq:15}c_n = 5.98 D_{\rm eff}(h_n)^{-1/3} C_n^2 dh / \cos(z) .%
\end{equation}%
Figure \ref{fig:2} shows the expected theoretical covariance functions given by Eqs. (\ref{eq:11}) and (\ref{eq:12}) for single seeing layers at different heights. We have made these calculations by assuming a 5.5~arcsec diameter for the FOV used by wavefront sensing. We assumed ten subapertures of 9.8~cm diameter each across the 98~cm SST pupil diameter, giving a total of 85
well illuminated subapertures within the pupil. The maximum field angle of 46.4~arcsec corresponds to the WFWFS built for the Swedish 1-m Solar Telescope. The calculations were made for heights $h=$ 0.0, 0.5, 1.5, 2.5, 3.5, 4.5, 6.0, 9.5, 16 and 30~km. From these covariance functions we conclude that the S-DIMM+ should be able to distinguish seeing at the pupil ($h=0$) and at a height of about 500~m above the pupil and that the angular resolution is adequate to allow measurements up to a height of about 20--30~km. However, the similarity between the covariance functions at 16 and 30~km height clearly demonstrates poor height resolution at these heights. This is a direct consequence of the large FOV used for wavefront sensing. As can be seen in the figure, the minimum height for which $r_0$ can be measured is set by the maximum field angle while the maximum height for which meaningful inversions can be made is set by the diameter of the FOV \citep{2002MNRAS.337..103W}. A particular feature in the covariance functions shown is the tilted dark line, marking minimum covariance. This corresponds to $s=\alpha h$, i.e., where the beam defined by $(s=0,\alpha)$ crosses the beam defined by $(s,\alpha=0)$ such that $\alpha=s/h$ --- see \refpic{fig:1}. This tilted dark line in principle allows direct measurement of the height of a single strong seeing layer, if that layer has a small vertical extent. 

\section{Implementation}
\subsection{Wavefront sensor description}
The wide-field wavefront sensor (WFWFS) is mounted immediately under the vacuum system of the SST. Light to the WFWFS is fed from a FOV adjacent to the science FOV and deflected horizontally such that \textit{the WFWFS beam does not pass the SST tip-tilt and AO system}. The WFWFS optics consists of a field stop, a collimator lens and an array with 85 hexagonal microlenses within the 98-cm pupil diameter, the layout of which is shown in \refpic{fig:3}. The microlenses have an equivalent diameter close to 9.8~cm. We record data through a 10~nm FWHM 500~nm CWL interference filter using a Roper Scientific 4020 CCD with $2048\times 2048$ $7.4\times 7.4 \mu$m pixels, operating at a frame rate of approximately 9~Hz and with an exposure time of 3.0~msec. The image scale is 0.344~arcsec per pixel.

\subsection{Image shift measurements}
\begin{figure}[t]%
   \centering
   \includegraphics[width=0.75\hsize]{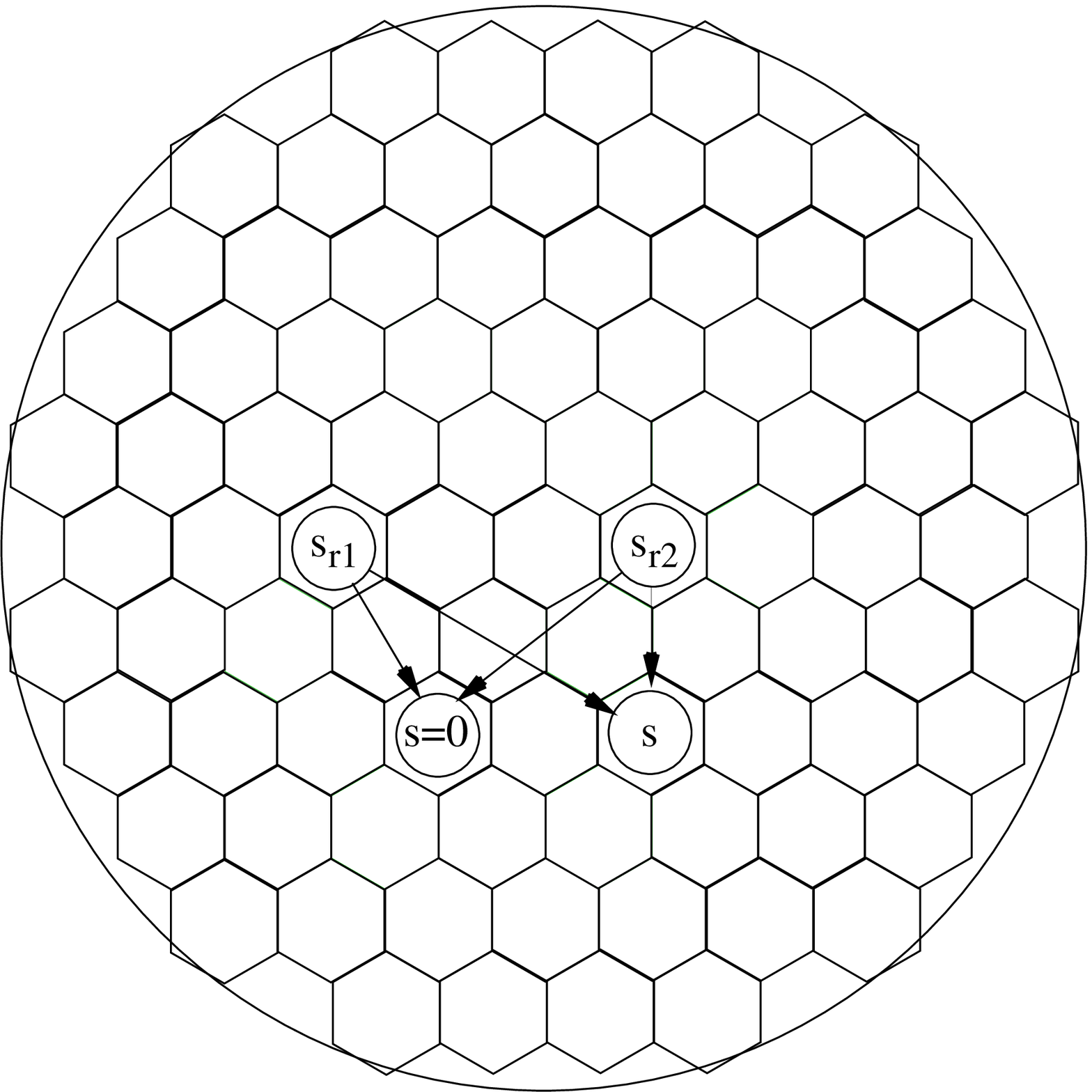}
   \vspace{2mm}
   \includegraphics[bb=285 0 1460 780, clip,width=0.75\hsize]{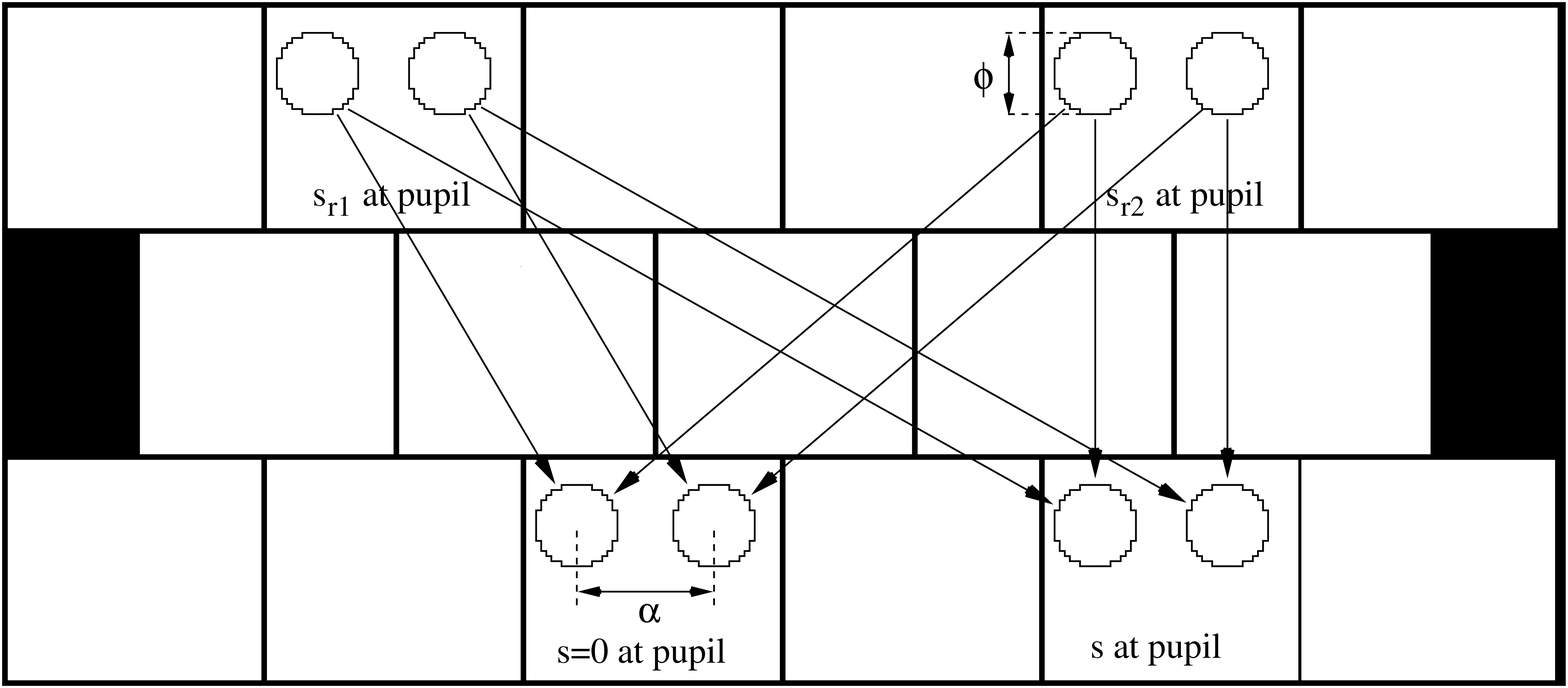}
   \vspace{2mm}
   \includegraphics[width=0.75\hsize]{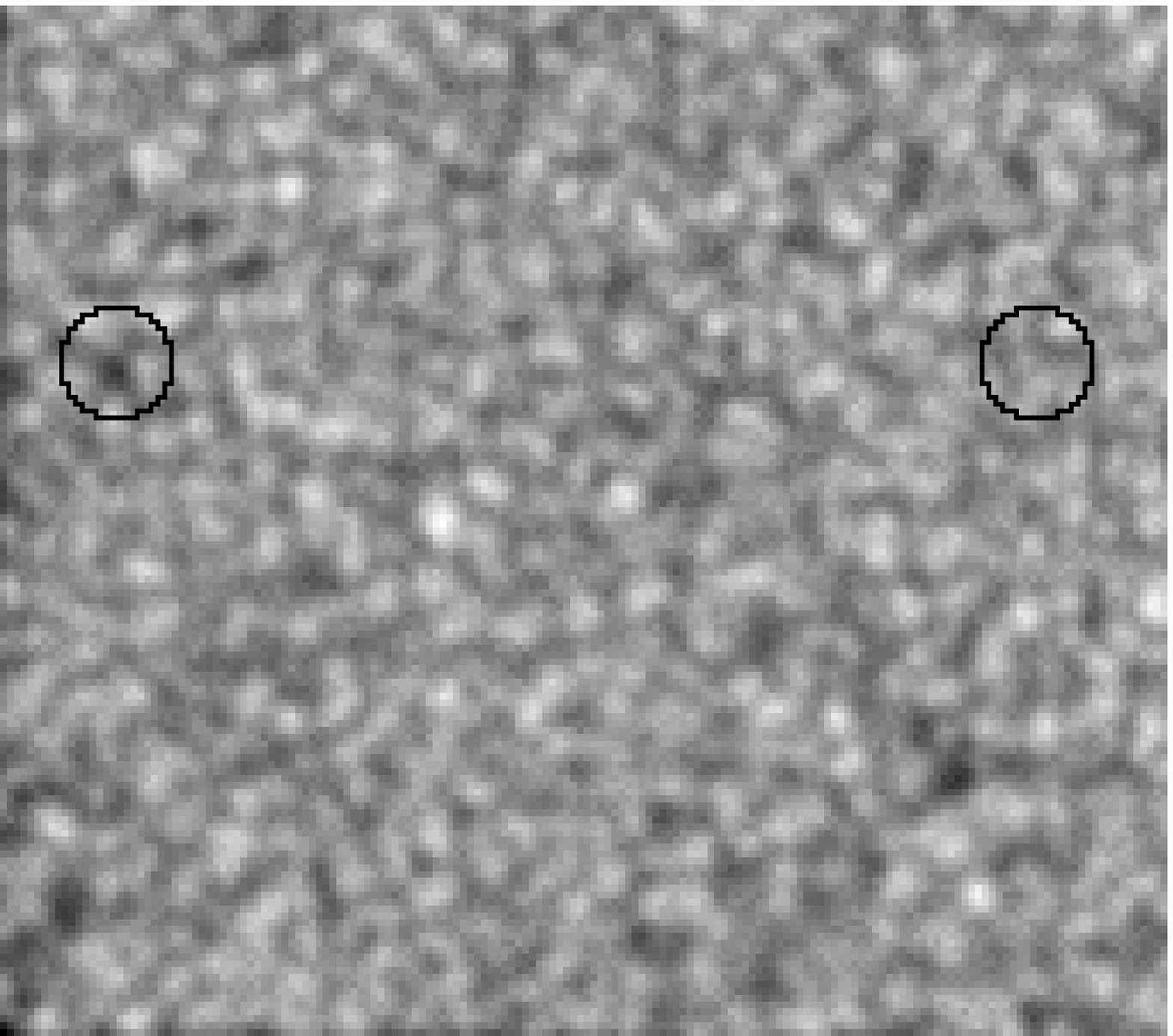}
      \caption{\label{fig:3}The upper figure shows the layout of the 85 fully illuminated hexagonal subapertures within the 98-cm SST aperture, indicated with the circle shown. Also indicated are two subapertures ($s_{r1}$ and $s_{r2}$), corresponding to selected {\it reference subimages} with high RMS contrast. Arrows point to two subapertures ($s=0$ and $s$) for which differential image shifts are measured with cross-correlation techniques. The mid panel shows the corresponding {\it subimages} with masks (not to scale) indicating two {\it subfields} at field angles separated by $\alpha$. The lowermost panel shows one of the 170x150 pixel subimages with granulation structure. The roundish masks outline the size and shape of the FOV actually used for differential image shift measurements.} 
\end{figure}%
The lower panel in \refpic{fig:3} shows a {\it subimage} with granulation, recorded with the WFWFS through one of its 85 subapertures. The subimage shown consists of 170x150 pixels, corresponding to 58.4x51.6~arcsec. In the left- and right-hand parts of that panel are indicated a roundish mask, outlining two {\em subfields} at different field angles $\alpha$, each with a diameter of approximately 16~pixels, or $\phi=5.5$~arcsec. The granulation pattern within a mask of this and another subimage from another subaperture is used to measure differential image shifts between two subapertures, as follows:

With the mask $M(x,y)$ at the first subimage fixed, the mask at the second subimage is moved by (m,n) pixels in the (x,y) image plane, while accumulating the squared intensity difference between the two images,
\begin{equation}%
\label{eq:16}\delta I^2(m,n) = \sum_{x,y} [M(x,y) I_1(x,y) - M(x+m,y+n) I_2(x,y)]^2 ,
\end{equation}%
where $I_1(x,y)$ and $I_2(x,y)$ correspond to the two granulation images and $M$ is unity within its 16~pixel diameter and zero outside. This defines $\delta I^2(m,n)$ at a number of discrete pixels surrounding the minimum corresponding to the best fit differential image shift. To improve the estimated image shift, the quadratic interpolation algorithm proposed by \citet[Eq. (10)]{1992lest.rept....1Y}, is used. Tests with this image shift measurement algorithm indicate that an RMS error of about 0.02--0.05~arcsec is achievable when $r_0$ is in the range 7-20~cm (L\"ofdahl, in preparation). Following accepted nomenclature, we refer to the procedure outlined above as a ``cross correlation'' technique and $\delta I^2(m,n)$ as ``correlation function'', although strictly speaking, we are not measuring image shifts with a cross correlation technique.

\subsection{Reference image frame selection and averaging} 
The low contrast (typically 3\% RMS in good seeing) granulation images observed through 9.8~cm subaperture and the relatively large (about 2~arcsec) granules, combined with the need to use a small FOV for wavefront sensing represents a significant challenge for wavefront sensing. Smearing of the granulation images by telescopic or atmospheric aberrations may cause large errors in the measured differential image shifts, or even complete failures. To reduce the risk of image shift measurements that fail completely, we do not perform image shift measurements directly between two arbitrary subimages. Instead, we use reference images, selected as the subimages with the highest RMS contrast. In \refpic{fig:3}, top panel, is indicated two subapertures (at locations $s_{r1}$ and $s_{r2}$) corresponding to such selected subimages and their relation to the two subapertures ($s=0$ and $s$) for which differential image shifts are measured. In the mid panel of \refpic{fig:3} the corresponding subimages and the {\em subfields} at two field angles, outlined with the roundish mask ($M$), are indicated schematically. The arrows in the top two panels of \refpic{fig:3} indicate how image shifts are measured. We first measure the differential image shift between reference subfield 1 (at $s_{r1}$) and the subfield at s=0 and also the image shift between reference subfield 1 and the subfield at $s$. We then calculate the difference between these two measured image shifts. We can express this operation as
\begin{multline}%
\label{eq:17}\delta x(s,\alpha)=[x_{r1}(s_{r1},\alpha)-x(0,\alpha)]_{cc}-[x_{r1}(s_{r1},\alpha)-x(s,\alpha)]_{cc} \\
  =x(s,\alpha)-x(0,\alpha) ,
\end{multline}%
where $x_{r1}(s_r,\alpha)$ is the position of the reference image measured with a subaperture at an (arbitrary) location $s_{r1}$ on the pupil and $[...]_{cc}$ indicates a measurement obtained with the cross-correlation technique outlined above. This cancels out the shift of the cross-correlation reference image, removes effects of any telescope movements (tip-tilt errors) and produces the {\em relative} image positions $(\delta x,\delta y)$ needed for the analysis. To further reduce noise, we repeat this process 4 times, using the best 4 subimages as references, and average the results of these 4 measurements. We emphasize that this corresponds to averaging 4 measurements of the {\em same} quantity and that this is done {\em separately} for each pair of subapertures and subfields at each field angle $\alpha$.  

\subsection{Zero-point references and time averaging}
The differential image shift measurements obtained with cross-correlation techniques suffer from a lack of absolute zero-point reference. Such an absolute reference is in practice impossible to define precisely and furthermore not needed. Instead, we rely on the differential image shifts to average to zero over a sufficiently long time interval. The WFWFS data collection system is set to record bursts of 1000~frames, corresponding to approximately 110~sec wavefront data. In the data reduction, we assume that the seeing induced differential image shift averaged over 110~sec is zero and subtract the average shift measured from the 1000~frames of the burst. {\em This averaging and bias subtraction is done separately for each pair of subfields ($\alpha$) within each pair of subimages at $s$ and $s=0$ and for each of the 4 cross correlation reference images.} This ensures that the measured covariances do not contain products of averages. Subtracting bias {\em individually} for data from the 4 cross correlation references is not needed for the covariance functions (we could just subtract the average shift for the image shifts averaged over the four cross correlation reference images), but is needed to correctly estimate the noise bias, as described in the following section.

\subsection{Noise bias estimation and compensation}
A fundamental limitation of this technique is the strong ground-layer seeing during day-time and the relatively weak contributions from higher layers. A weak seeing layer with $r_0=35$~cm contributes $(35/10)^{5/3}=8$ times less variance than a seeing layer with $r_0=10$~cm. The effective wavefront sensing FOV diameter $D_{\rm eff}$ is 80~cm at 30~km distance (Sect. 2). According to \refeq{eq:13}, the variance contributed from seeing at 30~km distance is thus reduced by approximately a factor $(0.8/0.1)^{1/3}=2$ by the averaging effect of the 5.5~arcsec FOV used to measure image displacements. Measurements of weak high-altitude seeing are therefore quite sensitive to random errors in the measured image positions, in the following referred to simply as ``noise''. In particular, this noise produces bias in the measured covariances that in turn leads to systematic errors, unless the noise bias is compensated for.

To reduce noise, we repeat each differential image shift measurement 4 times with 4 different cross-correlation reference images (Sect. 3.3). The corresponding image shifts are written for $i=$1--4 as
\begin{equation}%
\label{eq:18}\delta x_i(s,\alpha) = \delta x(s,\alpha)+ n_i(s,\alpha)) .
\end{equation}%
Here $\delta x(s,\alpha)$ is the true shift and $n_i(s,\alpha)$ is the noise of measurement $i$. The four measurements are averaged to reduce noise before calculating the covariance matrix. In addition, these four measurements are used to estimate and compensate for covariance noise \textit{bias}. The measured covariance $C_x^m$ in the presence of noise is
\begin{multline}%
\label{eq:19}C_x^m(s,\alpha) = \langle [\delta x(s,0)+ \overline n(s,0)] [\delta x(s,\alpha)+ \overline n(s,\alpha)] \rangle \\ = \langle \delta x(s,0) ~ \delta x(s,\alpha) \rangle + \langle \overline n(s,0) ~ \overline n(s,\alpha) \rangle\\= C_x(s,\alpha) + \delta C_x(s,\alpha).
\end{multline}%
where $\overline n(s,0)$ is the noise averaged from the four measurements,
\begin{equation}%
\label{eq:20}\overline n(s,\alpha)=\frac{1}{4} \sum_{i=1}^4  n_i(s,\alpha) ,
\end{equation}%
$C_x(s,\alpha)$ is the covariance matrix in the absence of noise and the covariance noise bias is
\begin{equation}%
\label{eq:21}\delta C_x(s,\alpha) = \langle \overline n(s,0) ~ \overline n(s,\alpha) \rangle .
\end{equation}%
To evaluate the expression for this noise bias, \textit{we assume that the noise of each measurement is uncorrelated with that of other measurements}, $\langle n_i(s,0)  ~ n_j(s,\alpha) \rangle = 0$, unless $i=j$. We then obtain
\begin{equation}%
\label{eq:22}\delta C_x(s,\alpha) = \frac{1}{16} \sum_{i=1}^4 \langle n_i(s,0) ~ n_i(s,\alpha) \rangle 
\end{equation}%
To estimate the noise bias directly from the data, we subtract the first measurement $\delta x_1(s,\alpha)$ from the corresponding average of all four measurements and calculate the covariance. We obtain the quantity
\begin{multline}%
\label{eq:23} D_1(s,\alpha) = \langle ( n_1(s,0) - \overline n(s,0) ) ( n_1(s,\alpha) - \overline n(s,\alpha) ) \rangle \\ = \left< [ (\frac{3}{4} n_1(s,0) - \frac{1}{4} \sum_{i=2}^4 n_i(s,0) ] ~ [ (\frac{3}{4} n_1(s,\alpha) - \frac{1}{4} \sum_{i=2}^4 n_i(s,\alpha) ]  \right> .
\end{multline}%
Evaluating this expression, we obtain
\begin{multline}%
\label{eq:24}D_1(s,\alpha) = \frac{9}{16} \langle n_1(s,0) ~ n_1(s,\alpha) \rangle + \frac{1}{16} \sum_{i=2}^4 \langle n_i(s,0) ~ n_i(s,\alpha) \rangle  .
\end{multline}%
Forming similar expressions $D_i(s,\alpha)$, with i=2,3,4, for the remaining three measurements and adding these, we obtain
\begin{equation}%
\label{eq:25}\sum_{i=1}^4 D_i(s,\alpha) = \frac{12}{16} \sum_{i=1}^4 \langle n_i(s,0) ~ n_i(s,\alpha) \rangle .
\end{equation}%
Comparing to \refeq{eq:22}, this gives the desired estimate of the noise bias $\delta C_x(s,\alpha)$ as 
\begin{equation}%
\label{eq:26}\delta C_x(s,\alpha) = \frac{1}{12} \sum_{i=1}^4 D_i(s,\alpha) .
\end{equation}%
\textit{This noise bias is estimated at each $s,\alpha$ individually, for each data set separately} and subtracted from the measured covariance matrix $C_x^m$.

We note that \refeq{eq:23} is insensitive to any {\it random} errors that are {\it common} to (the same for) all 4 measurements. Thus, the proposed noise estimation method is useful to indicate the magnitude of random errors in the measured image shifts, in particular in bad seeing, but is not sufficiently accurate to provide robust estimates of the noise bias. Simulations are needed for proper understanding of noise propagation effects.

\subsection{Least-squares solution method}
To obtain the unknown coefficients $c_n$ in Eqs. (\ref{eq:9}) and (\ref{eq:10}), we can solve a conventional linear least-squares fit problem by minimizing the badness parameter $L$, given by
\begin{multline}%
\label{eq:27}L = \sum\limits_{s,\alpha} \left( [C_x(s,\alpha) - \sum\limits_{n=1}^{N} c_n F_x(s,\alpha, h_n)]^2 \right.   \\ + \left. [C_y(s,\alpha) - \sum\limits_{n=1}^{N} c_n F_y(s,\alpha, h_n)]^2 \right) W^2(s,\alpha) , 
\end{multline}%
with respect to the coefficients $c_n$. Here, the weight $W(s,\alpha)$ equals the number of independent measurements for each $(s,\alpha)$ (\refpic{fig:2}, lowermost panel). As is evident from the layout of the microlens array (\refpic{fig:3}), a large number of independent samples is possible for small separations $s$, but only a few samples are possible for large $s$. Similarly, a large number of samples are possible for small, but not large, field angles $\alpha$. The weights $W(s,\alpha)$ in \refeq{eq:27} properly balance the least-squares fit to take into account the strongly varying number of measurements for different $s,\alpha$ and the corresponding noise. This is of particular importance for (weak) high-altitude seeing, the signatures of which are at small angles $\alpha$, where $W$ is large, 

Minimizing $L$ leads to a linear matrix equation $\bf{A c = b}$ for $c_n$. However, this permits solutions with negative values for $c_n$, which is clearly not physical. In order to restrict the solutions to yield positive values for $c_n$, we make the variable substitution 
\begin{equation}%
\label{eq:28}c_n = \exp(y_n) ,
\end{equation}%
(Collados, private communication) and solve the corresponding \textit{non-linear} least-squares fit problem with respect to the parameters $y_n$. Fits to data obtained so far indicates excellent convergence properties of the implemented non-linear method. 

\subsection{Height grid optimization}

\begin{figure}[t]%
   \centering
   \includegraphics[bb=45 35 570 740, clip, width=0.7\hsize,angle=-90]{13791fg4a.eps}
   \includegraphics[bb=45 35 570 740, clip, width=0.7\hsize,angle=-90]{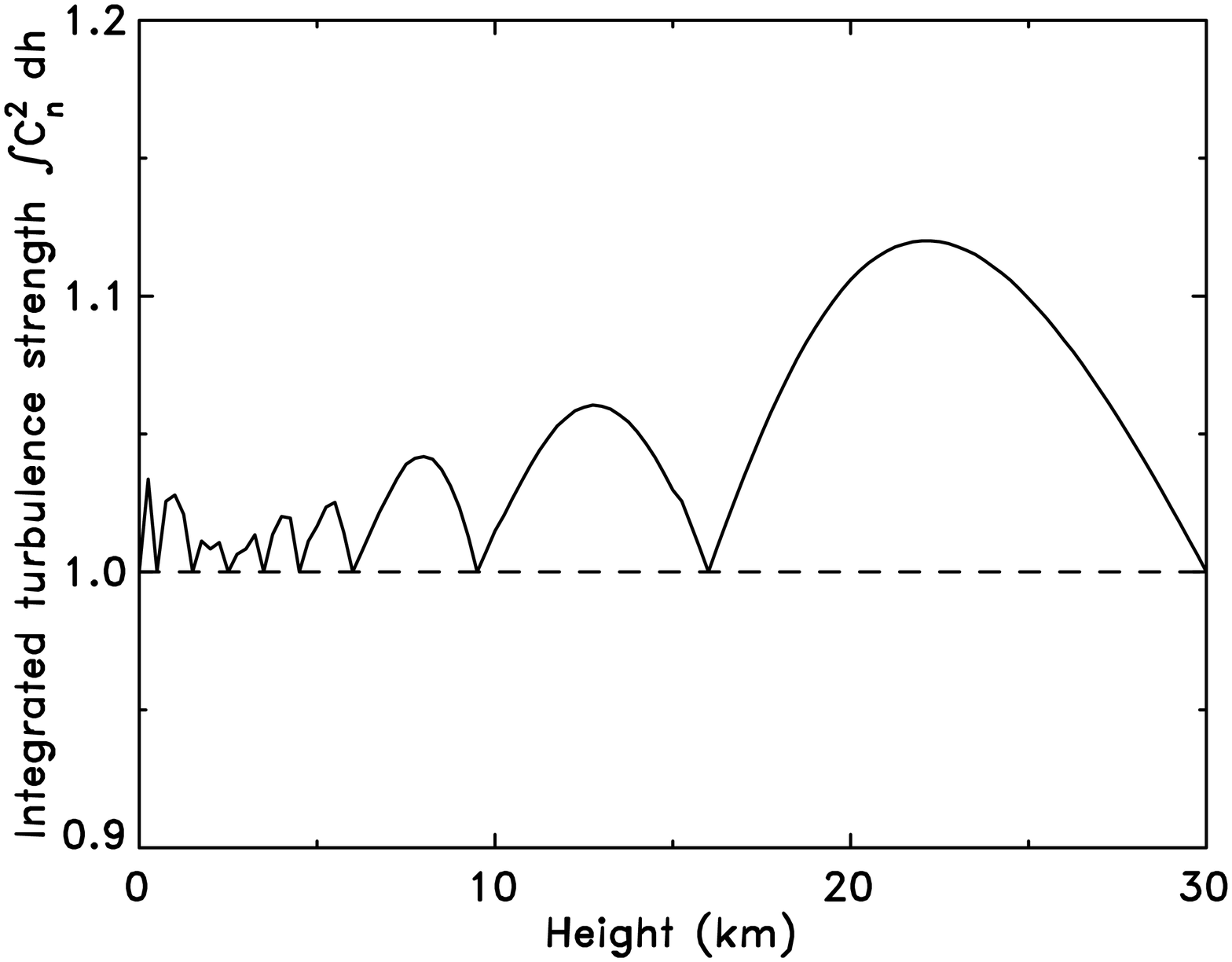}
      \caption{\label{fig:4}Response of inversion code to a single thin seeing layer of unit strength ($c_n=1$) located at variable height (top panel). The vertical bars represent the $c_n$ values at different heights (nodes), returned by the inversion code. The lower panel shows the returned integrated turbulence strength ($\sum c_n$) at each height.}
\end{figure}

Good height grids can be found by calculating the inverse of the matrix $\bf A$, corresponding to the linear solution for $c_n$, and choosing a height grid
that minimizes its noise sensitivity (sum of squared elements of the inverse of $\bf A$). Such optimizations show that we should be able to determine contributions from the pupil plane plus about 8--9 layers above the pupil with the lowermost layer above the pupil located at a height of 500~m. The maximum height can be 30~km with a FOV of $5.5 \times 5.5$~arcsec, however the height below that must be located around 16~km to not cause high noise amplification. The height resolution with which we can determine seeing contributions from layers above 10~km with this large FOV is thus strongly limited. Only with a smaller FOV can the height resolution at large distances be improved. For the inversions discussed in this paper, we used the height grid defined by $h$= 0.0, 0.5, 1.5, 2.5, 3.5, 4.5, 6.0, 9.5, 16 and 30~km. We tested this configuration with input seeing layers at heights in steps of 250~m from 0 to 30~km height, then solved for contributions from the 10-layer height grid defined above. For input heights that matched one of the 10-layer heights, the inversion recovers the input height and contribution perfectly. When the input height is in between two of the heights in the inversion model, the inversion responds by distributing the correct $c_n$'s between the two surrounding layers such that the relative distributions are in rough proportion to the difference between the true height and the two surrounding heights in the inversion model. By constraining the $c_n$ values to be positive, negative overshoot in adjacent layers is prevented. This is illustrated in \refpic{fig:4}, which shows the response of the inversion to a thin seeing layer located at variable height\footnote{ This plot is similar to that showing the height response for the multi-aperture scintillation sensor (MASS) \citep{2003MNRAS.343..891T}.}. Due to the coarse height grid in the upper layers, the integrated turbulence strength is overestimated by up to 12\%. A denser grid would reduce that overestimate but also increase noise amplification.  
We conclude that the method should work well with good input data.
\begin{figure}[]%
   \centering
   \includegraphics[bb=53 60 570 705, clip, width=0.7\hsize,angle=-90]{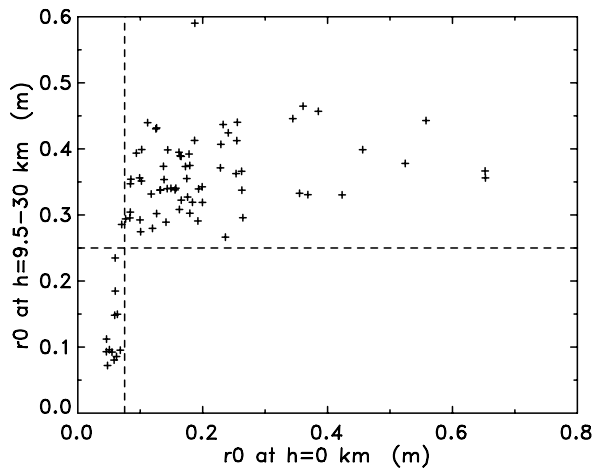}
   \caption{\label{fig:5}The figure shows $r_0$ for the ground-layer ($h=0$) versus the integrated $r_0$ for the 9.5, 16 and 30~km layers and demonstrates that all inversions that return small values for $r_0$ at the highest layers
    correspond to poor ground-layer seeing (small $r_0$ at $h=0$).}
\end{figure}%

\begin{figure}[t]%
   \includegraphics[width=0.45\hsize]{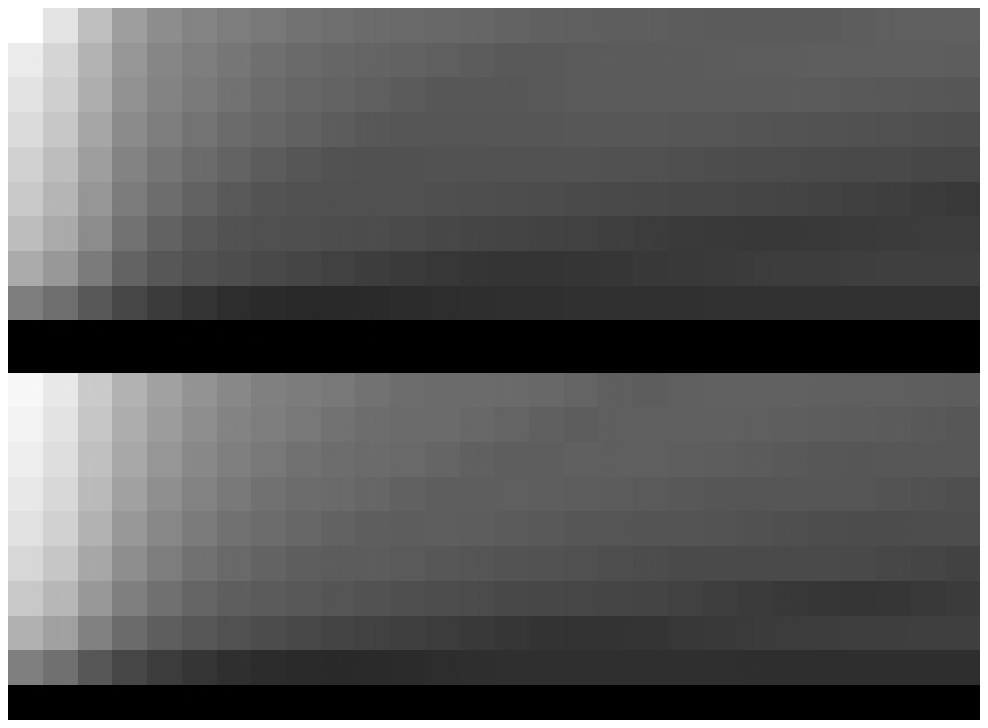}
   \includegraphics[width=0.45\hsize]{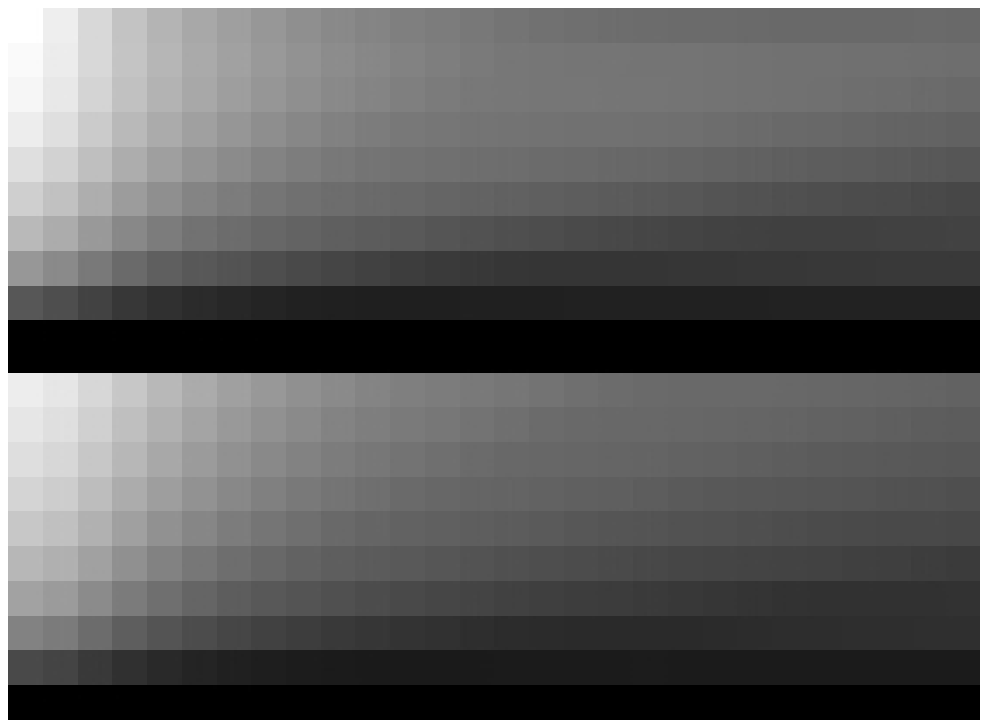}\\[0.9mm]
   \includegraphics[width=0.45\hsize]{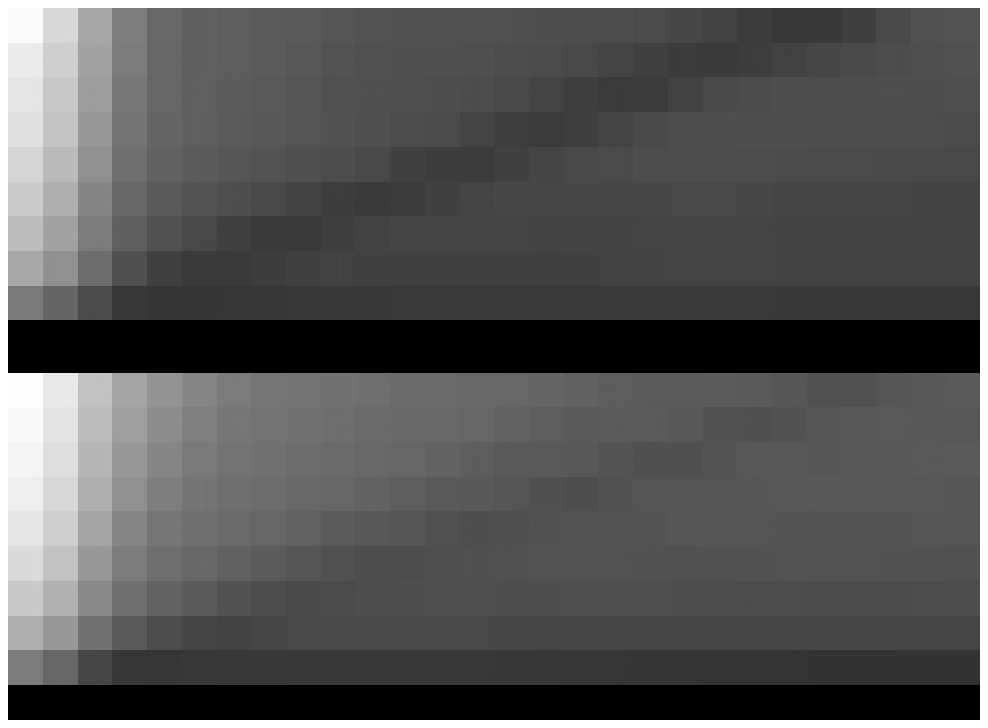}
   \includegraphics[width=0.45\hsize]{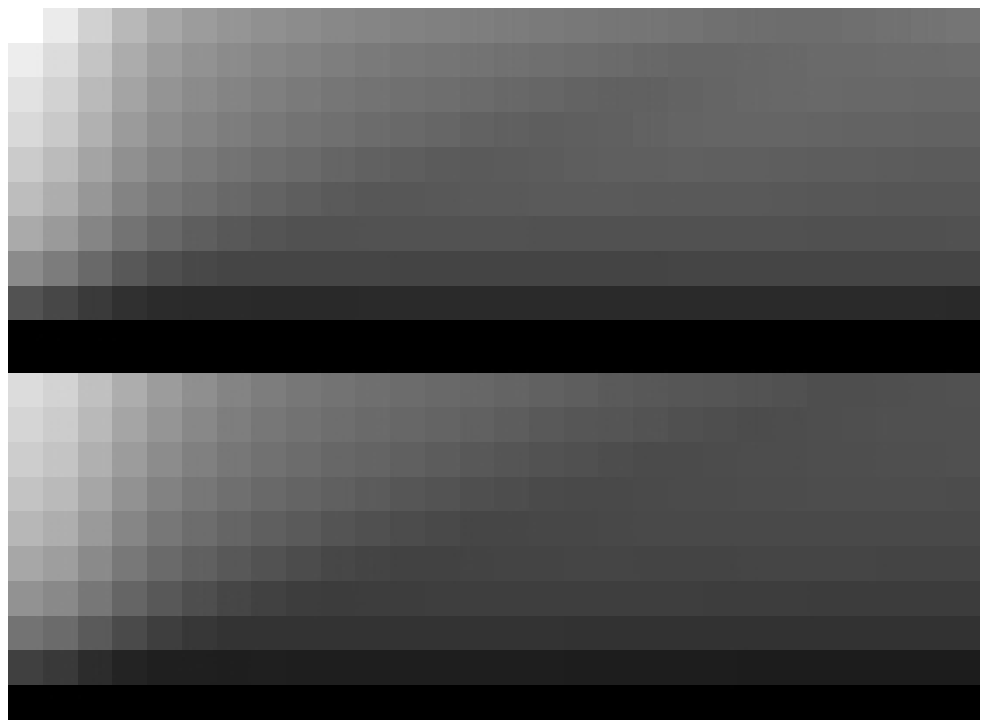}\\[0.9mm]
   \includegraphics[width=0.45\hsize]{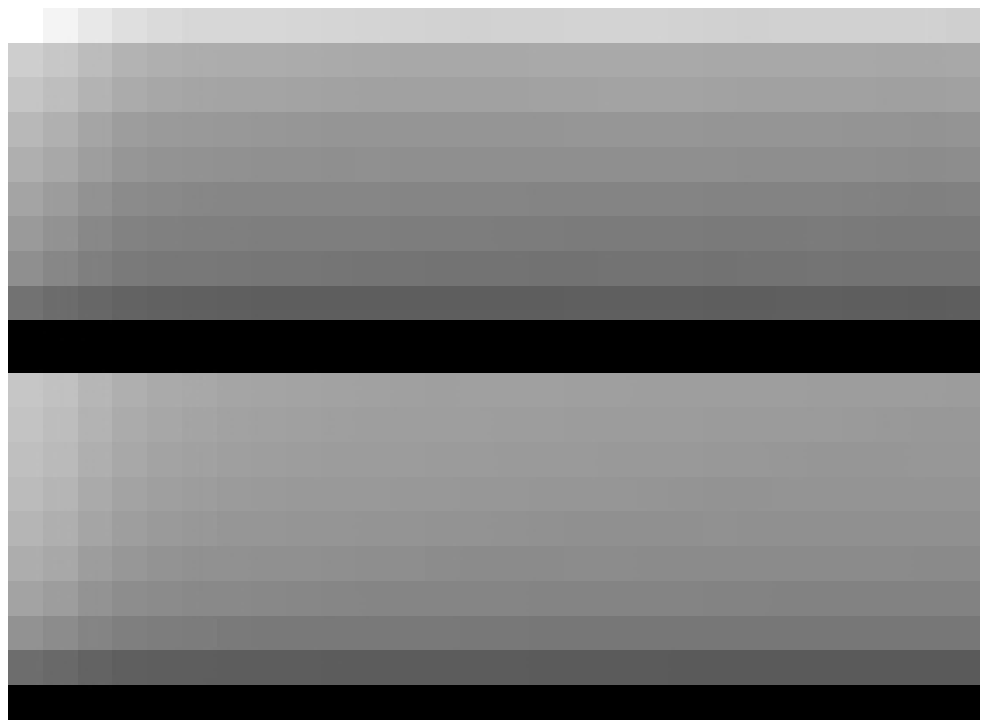}
   \includegraphics[width=0.45\hsize]{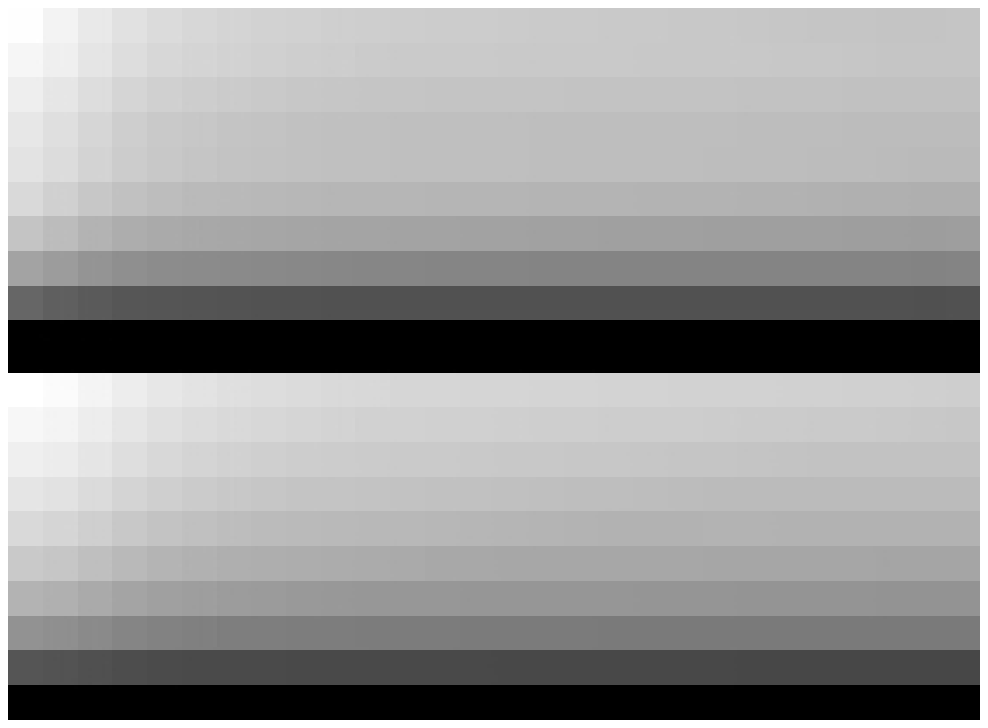}\\[0.9mm]
   \includegraphics[width=0.45\hsize]{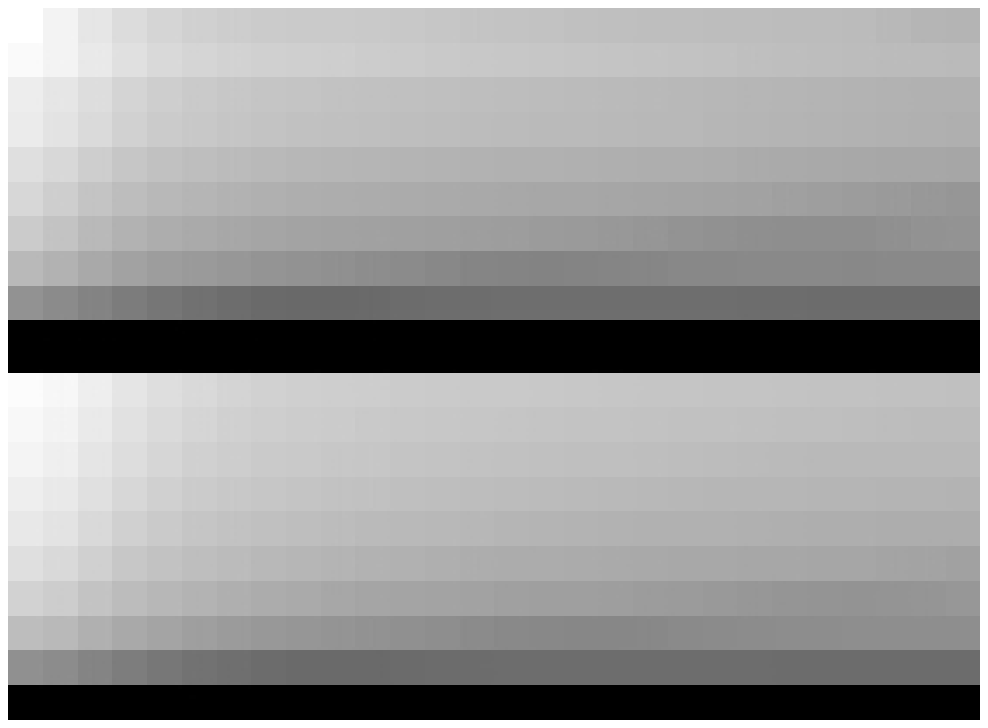}
   \includegraphics[width=0.45\hsize]{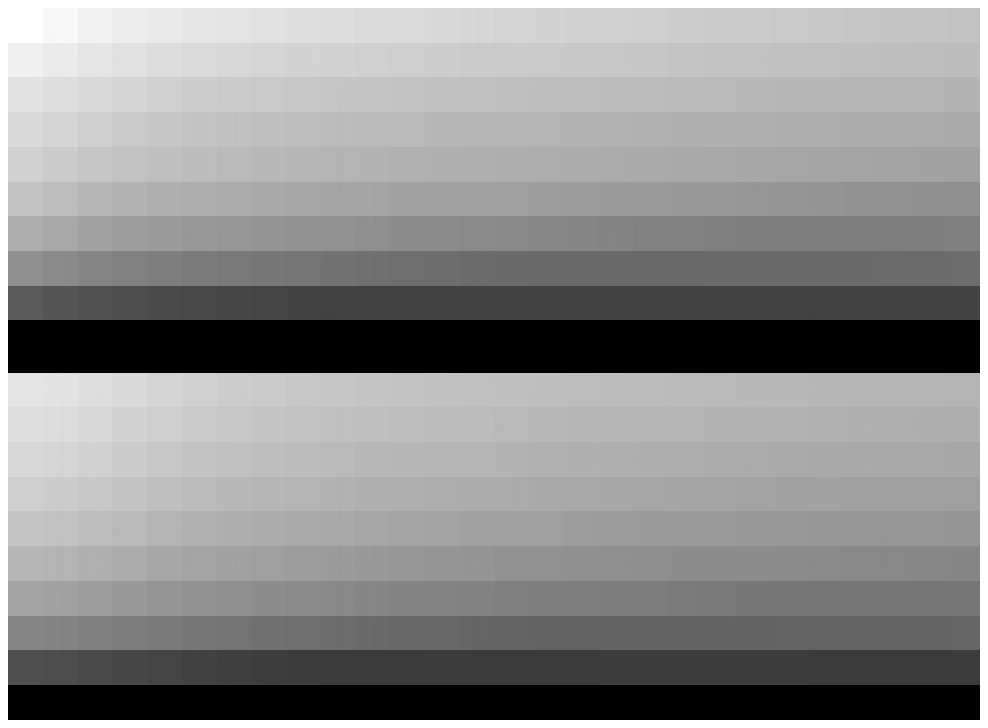}\\[0.9mm]
   \includegraphics[width=0.45\hsize]{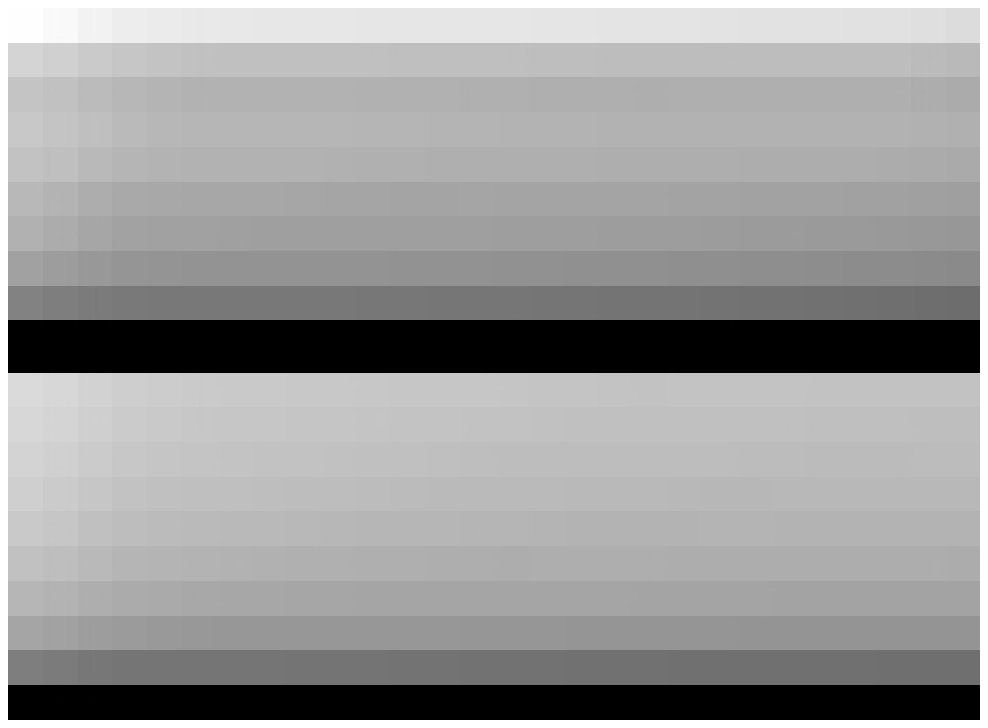}
   \includegraphics[width=0.45\hsize]{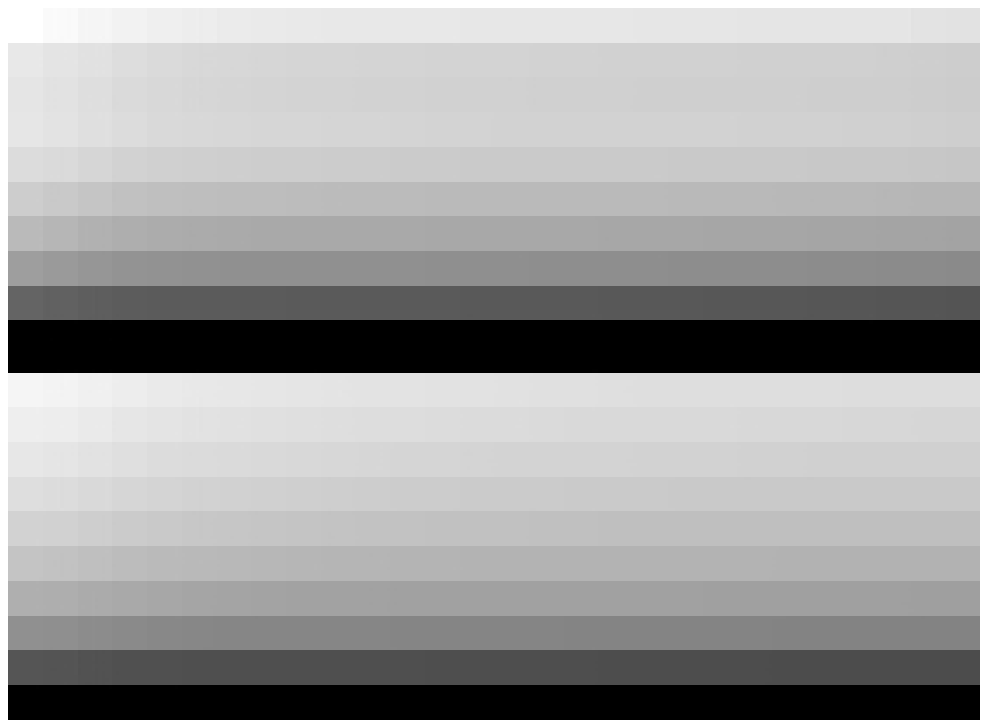}\\
   \caption{\label{fig:6}Covariance functions for 5 data sets processed (00:1, 03:0, 07:3, 09:3 and 15:3). Within each of the panels is shown the observed (top) and modeled (bottom) covariance. Longitudinal covariances are shown in the left column, transverse covariances in the right column. The separation between the subapertures $s$ increases upwards and the field angle $\alpha$ increases to the right in each sub-panel. Compare to the theoretical covariance functions for layers at different heights in \refpic{fig:2}. Data and fits shown are based on covariance functions measured along rows of microlenses and sub-fields only.}
\end{figure}%

\begin{figure}[]%
   \centering
   \includegraphics[bb=53 50 570 705, clip, width=0.38\hsize,angle=-90]{13791fg7a.eps}
   \includegraphics[bb=53 50 570 705, clip, width=0.38\hsize,angle=-90]{13791fg7b.eps}
   \includegraphics[bb=53 50 570 705, clip, width=0.38\hsize,angle=-90]{13791fg7c.eps}
   \includegraphics[bb=53 50 570 705, clip, width=0.38\hsize,angle=-90]{13791fg7d.eps}
   \includegraphics[bb=53 50 570 705, clip, width=0.38\hsize,angle=-90]{13791fg7e.eps}
   \includegraphics[bb=53 50 570 705, clip, width=0.38\hsize,angle=-90]{13791fg7f.eps}
   \includegraphics[bb=53 50 570 705, clip, width=0.38\hsize,angle=-90]{13791fg7g.eps}
   \includegraphics[bb=53 50 570 705, clip, width=0.38\hsize,angle=-90]{13791fg7h.eps}
   \includegraphics[bb=53 50 570 705, clip, width=0.38\hsize,angle=-90]{13791fg7i.eps}
   \includegraphics[bb=53 50 570 705, clip, width=0.38\hsize,angle=-90]{13791fg7j.eps}
   \caption{\label{fig:7}Plots of measured (solid) and modeled (dashed) longitudinal (left column) and transverse (right column) covariance as function of pupil separations $s$ at field angles $\alpha=0$~arcsec (upper curves) and $\alpha=36.1$~arcsec (lower curves). Plots for data sets in \refpic{fig:6} (00:1, 03:0, 07:3, 09:3 and 15:3, top to bottom) are shown, using only covariances calculated along rows.}
\end{figure}%
\begin{figure}[]%
   \centering
   \includegraphics[bb=53 50 570 705, clip, width=0.38\hsize,angle=-90]{13791fg8a.eps}
   \includegraphics[bb=53 50 570 705, clip, width=0.38\hsize,angle=-90]{13791fg8b.eps}
   \includegraphics[bb=53 50 570 705, clip, width=0.38\hsize,angle=-90]{13791fg8c.eps}
   \includegraphics[bb=53 50 570 705, clip, width=0.38\hsize,angle=-90]{13791fg8d.eps}
   \caption{\label{fig:8}Plots of measured (full) and modeled (dashed) longitudinal and transverse covariance as function of pupil separations $s$ at field angles $\alpha=0$~arcsec (upper curves) and $\alpha=36.1$~arcsec (lower curves). Plots for data sets 03:0 and 09:3 are shown (top to bottom), combining covariances measured along both rows and columns. Compare to \refpic{fig:7}.}
\end{figure}%
\begin{figure}%
\centering
    \includegraphics[bb=55 55 570 705, clip, width=0.38\hsize,angle=-90]{13791fg9a.eps}
    \includegraphics[bb=55 65 570 705, clip, width=0.38\hsize,angle=-90]{13791fg9b.eps}
    \includegraphics[bb=55 65 570 705, clip, width=0.38\hsize,angle=-90]{13791fg9c.eps}
    \caption{\label{fig:9}Noise covariance and related data. The first plot shows the measured covariance (upper) and 10 times the measured noise covariance (lower) at $s=0.098$~m as function of field angle $\alpha$ for data set 03:0. This demonstrates the rapid drop (and reduced impact) of the covariance noise with field angle. The second plot shows the relation between the measured ground layer seeing ($r_0$ at $h=0$) and the noise in the measured image positions (square root of the noise covariance). The vertical dashed line corresponds to $r_0=7.5$~cm at $h=0$, confirming our previously determined threshold for useful data. The third plot shows that the noise in the measured positions increases with anisoplanatism, here quantified as the combined $r_0$ for the two highest (16--30~km) layers used in the inversions.}
\end{figure}%

\section{First Results}
We recorded 20 bursts of wavefront data, each consisting of 1000 exposed frames, between 8:40 and 15:00 UTC on June 26, 2009. Each burst of 1000 frames took approximately 110~sec to record. After estimating and removing noise bias from each 1000 frame burst, the noise compensated data was analyzed in blocks of 250 frames, corresponding to about 27~sec of seeing data. While recording this wavefront data (without AO), the SST was used to record science data with its AO system in closed loop. The initial seeing quality was considered good. Around 11 UTC, the seeing started to deteriorate rapidly and science observations were stopped at 11:03 UTC.

Due to problems with data acquisition, several data sets (16--19) contained corrupt images and were not processed. 

A fundamental aspect of the system is under what seeing conditions reliable inversions are possible. Several inversions return very low values for $r_0$ in the uppermost layers. Figure \ref{fig:5} shows $r_0$ for the ground-layer ($h=0$) versus the integrated $r_0$ for the 9.5--30~km layers. The dashed vertical line corresponds to $r_0=7.5$~cm for $h=0$, the dashed horizontal line to $r_0=25$~cm at the 9.5--30~km layers. This figure shows that the smallest values for $r_0$ at the highest layers are all associated with poor ground-layer seeing. We conclude that inversions that return $r_0$ values at $h=0$ smaller than 7.5~cm, or about 75\% of the subaperture diameter, should be rejected and excluded these data sets from further analysis.

We now take a closer look at the data and the inversions. The covariance functions, defined in \refeq{eq:3}, were evaluated at steps of 5 pixels (1.72~arcsec) with the maximum field angle limited to 46.4~arcsec. Noise bias was subtracted from the data, as described in Sect. 3.5. The so-obtained covariance functions for 5 data sets recorded between 08:40 and 11:01 UTC (zenith distance in the range 30--61.0~deg) are shown in \refpic{fig:6} together with the modeled covariance functions. The data and fits shown in this figure and in \refpic{fig:7} are based on covariance functions calculated along rows of microlenses and corresponding sub-fields only (see \refpic{fig:3}), in \refpic{fig:8} are shown plots of observed and fitted data based on covariance functions calculated along both rows and columns of microlenses and sub-fields. The height grid used with these inversions has nodes at 0.0, 0.5, 1.5, 2.5, 3.5, 4.5, 6.0, 9.5, 16 and 30~km. Comparing \refpic{fig:6} to \refpic{fig:2}, the top two measured covariance functions indicate clear signatures of a ground layer plus one dominating high-altitude layer (note the slanted dark line in these panels). In \refpic{fig:7} are shown plots of these measured and modeled covariance functions as functions of $s$ for field angles $\alpha$ of 0 and 36.1~arcsec, where the second field angle is large enough to give very small influence from high-altitude seeing. It is evident that the fits are in general good, but that the measured transverse covariance is often too strong relative to that of the longitudinal covariance.

The jagged appearance of the plots in \refpic{fig:8} are due to differences in measured covariances from rows and columns. This appears to indicate turbulence that is not isotropic above the ground layer. We have made inversions with and without the column covariances included and established that the overall conclusions drawn in this paper are robust, even though systematic differences between the two types of inversions exist.
\begin{table}[tbh]%
  \centering
  \begin{tabular}{rrrrrrrrrrrrr}
    \hline
    \mathstrut
    &$\alpha=0$&$\alpha=0$&$\alpha=1.72$&$\alpha=1.72$&$\alpha>\phi$&$\alpha>\phi$&  \\
    \hline
Set&Long&Transv&Long&Transv&Long&Transv \\
    \hline
00:1&    25&    18&   156&   119&  8213&  4389 \\
03:0&    26&    18&   137&   102&  4375&  3058 \\
04:2&     4&     3&    25&    20&  3192&  5242 \\
07:3&    28&    19&   163&   116&  3541&  1315 \\
09:3&    24&    19&   131&    97&  1642&   650 \\
15:3&    77&    58&   310&   317&   927&  2426 \\
20:0&     2&     2&     5&     5&    11&     9 \\
22:1&    58&    42&   286&   199&  1347&   795 \\
\hline
  \end{tabular}
  \caption{\label{tab:1}Signal to noise ratio for covariance functions measured at $\alpha=0$, $\alpha=1.72$~arcsec and for de-correlated sub-fields ($\alpha > \phi$).} 
  \label{table_exp1}
\end{table}%

\subsection{Noise measurements}

In \refpic{fig:9} (left panel) is shown the variation of the covariance (upper curve) and noise covariance (lower curve, multiplied by a factor 10) with $\alpha$, at a separation $s=0.098$~m for data set 3:0. Note in particular, that the noise covariance at $s=0$ is about 6\% of the measured variance at $\alpha=0$, which is similar to the expected contribution from a seeing layer with $r_0=35$~cm at $h=30$~km when $r_0=10$~cm at $h=0$ (see Sect. 3.5). Note also that the noise covariance drops off with angle in a way that is reminiscent of high-altitude seeing. However, the noise covariance drops by a factor of nearly 8 at $\alpha=1.7$~arcsec, such that the noise covariance is negligible already at this field angle separation. This rapid decorrelation of the noise occurs on a scale that is similar to the diameter of individual granules, rather than on a scale that corresponds to the diameter of the FOV. Table 2 summarizes the S/N, defined as the ratio of the measured covariance (corrected for noise bias) to the noise covariance, for a few selected data sets. In this Table, covariances used to calculate S/N have been averaged over all $s$ in the range 0.098--0.784~m. The small S/N for data set 04:2 primarily comes from the excellent ground-layer seeing ($r_0=42$~cm) for this data set. Data set 20:0 corresponds to one of the data sets rejected due to bad ground-layer seeing. Data set 22:1 corresponds to relatively poor seeing, $r_0=7.7$~cm for $h=0$ and $r_0=6.9$~cm integrated over the atmosphere, but evidently the wavefront data correspond to excellent S/N at all angles $\alpha$.

The mid panel of \refpic{fig:9} shows the variation of the image position noise, calculated as the square root of the noise covariance at $s=0.098$~m and
$\alpha=0$, with $r_0$ at $h=0$. Quite clearly, the image position noise
increases dramatically when $r_0$ for the ground layer is less than about
7~cm. This renders wavefront sensor measurements essentially useless when
$r_0<7.5$~cm, as concluded already from \refpic{fig:5}. Finally, the last panel in \refpic{fig:9} shows a relation between the image position noise and $r_0$ at the highest layers, clearly demonstrating that anisoplanatism from high-altitude seeing causes problems with the large FOV used for wavefront sensing. When $r_0$ is larger than about 45~cm, the RMS noise is about 0.05~arcsec but when $r_0$ is 35~cm, the RMS noise is doubled. Since our FOV corresponds to a diameter of 40~cm at 16~km, where most of the high-layer seeing originates, our results confirm the expectation that the FOV should be smaller than $r_0$ at the height for which $r_0$ is determined, else wavefront sensor noise increases rapidly with decreasing $r_0$.

\begin{figure*}[]%
    \centering
    \includegraphics[bb=55 50 570 705, clip, width=0.26\hsize,angle=-90]{13791fg10a.eps}
    \includegraphics[bb=55 50 570 705, clip, width=0.26\hsize,angle=-90]{13791fg10b.eps}
    \includegraphics[bb=55 50 570 705, clip, width=0.26\hsize,angle=-90]{13791fg10c.eps}
    \includegraphics[bb=55 50 570 705, clip, width=0.26\hsize,angle=-90]{13791fg10d.eps}
    \includegraphics[bb=55 50 570 705, clip, width=0.26\hsize,angle=-90]{13791fg10e.eps}
    \includegraphics[bb=55 50 570 705, clip, width=0.26\hsize,angle=-90]{13791fg10f.eps}
    \caption{\label{fig:10}Summary of results. The variations of the turbulence strength $C_n^2 dh$ with time are shown at different heights. From left to right: Turbulence strength at 0, 0.5 and 1.5~km height (top row) and at 16 and 30~km (bottom row). The lower-right plot shows $C_n^2$ integrated over the 9.5--30~km layers. Note the different
    scales on the y-axis for these plots.}
\end{figure*}%

We finally caution that noise bias subtraction has a significant effect on measurements of seeing for the {\it highest} layers and that our confidence in these estimates relies on the method for noise compensation outlined in Sect. 3.5. It is therefore of considerable importance to study noise estimation by means of simulations.

\subsection{Discussion of results}

\begin{figure}[]%
   \centering
    \includegraphics[bb=55 50 580 725, clip, 
    width=0.74\hsize,angle=-90]{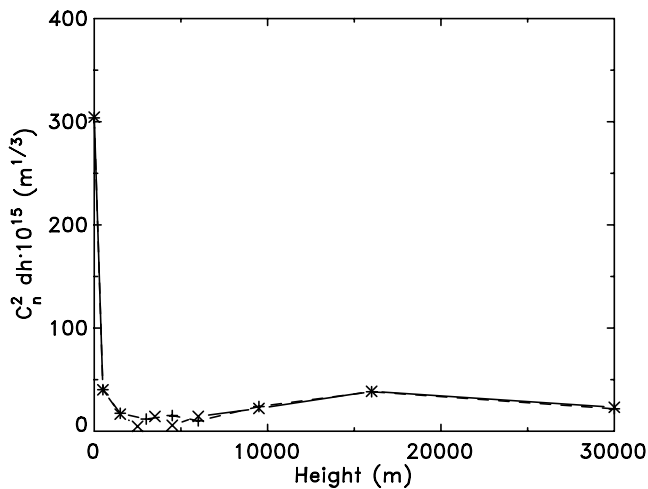}
    \caption{\label{fig:11}Summary of results. Shown is $C_n^2 dh$ at different heights, averaged over all measurements from June 26, 2009 (solid curve). Also shown (dashed curve) is the average results for inversions made with the nodes at 2.5 and 3.5~km replaced with a single node at 3~km. Note that the varying zenith distance is not compensated for in the averages.}
\end{figure}

Figures \ref{fig:10} and \ref{fig:11} summarize some of the results. These two figures are based on fits to measured covariances along both horizontal rows and vertical columns, corrected for noise bias (Sect 3.5). 

The top row of \refpic{fig:10} shows the variation of the turbulence strength $C_n^2 dh$ with time for the three near-ground layers ($h=0$, 0.5 and 1.5~km). The dominant layer is obviously at $h=0$, and its variation with time clearly indicates gradually degrading seeing. The large scatter is real and illustrates the intermittent nature of the seeing, when averaged over relatively short time intervals (here, 27~sec), for these particular observations. The seeing at 500~m altitude shows a similar trend of degrading with time, but with turbulence strength that is typically 8 times weaker than those at $h=0$. The seeing layer at 1.5~km is even weaker than that at 0.5~km. 

The time evolution of seeing at the two highest layers are shown in the lower row of plots in \refpic{fig:10}. Except for a few data sets, the seeing contributions are consistently small from the 30~km layer. During this period, the high-altitude seeing primarily comes from the 16~km layer. The combined $r_0$ for the 9.5, 16 and 30~km layers averages at 37~cm. The accuracy of these estimates of high-altitude seeing is difficult to assess without independent verification. Images recorded with the SST (diameter 98~cm) generally show small-scale geometrical distortion but only minor small-scale differential blurring over the FOV, except when the Sun is at large zenith distance or when observations are made at short wavelengths around 400~nm. This is consistent with $r_0$ values of about one third of the SST diameter or larger, in agreement with the values found from this data. As regards the actual height of this seeing layer, upper-air sounding data above Tenerife at 0:00 and 12:00 UTC for this day (\url{http://weather.uwyo.edu/upperair/sounding.html}, Guimar-Tenerife) show a temperature rise above 17--18~km altitude, indicating the location of the tropopause, and enhanced wind speeds between roughly 10.5 and 15.5~km, peaking at 14~km, altitude. If the latter layer is where the high-altitude seeing originates and if this is at the same height above La Palma as above Tenerife, then the height of that seeing layer should be about 11.5~km above the telescope. This corresponds to a distance of 23~km for the first data sets recorded at a zenith distance of 60~deg. This is midways between the two uppermost nodes in our inversion model.

However, the seeing estimates at 30~km are very uncertain and our confidence in these estimates rely on the applicability of the weights $W(s,\alpha)$, defined in Sect. 3, and on the noise bias estimation and subtraction method outlined in Sect. 3.1. Setting all weights $W(s,\alpha)$ equal to unity reduces the average estimates of $r_0$ at 30~km by about a factor of two, but has only a minor effect on $r_0$ at all other heights. Similarly, reducing the weight $W$ at $\alpha=0$ by a factor of two also strongly increases the estimated turbulence strength at 30~km but has a small effect at other heights. These and other tests indicate uncomfortably large uncertainties of our estimates of seeing at 30~km distance, most likely primarily related to the large FOV used.
 
An indication, based on SST observations, of seeing contributions from \textit{intermediate} heights is the absence of noticeable {\em large-scale} variations of image quality over a science FOV of typically 1~arcmin. This suggests that the dominant seeing is close to the ground layer and at high altitude and that $r_0$ is significantly larger than 30~cm for \textit{intermediate} layers. Our data are consistent with this. The solid curve in \refpic{fig:11} shows the turbulence strength $C_n^2 dh$ as function of height, averaged over all data from 26 June 2009. Contributions to the seeing from heights in the range 1.5--6~km are obviously small for this data set. A conspicuous feature in \refpic{fig:11} is the increased turbulence strength at 3.5~km, suggesting a weak seeing layer. This feature may not be real, but an artifact of too dense grid points in this height range. We repeated the inversions after replacing the two nodes at 2.5 and 3.5~km with a single node at 3~km, but leaving the remaining nodes unchanged. The result is indicated with a dashed line in \refpic{fig:11} and shows a smoother variation of $C_n^2 dh$ with height. 

\subsection{Extension of the method}

In the present paper, we have calculated covariance functions from measured $x,y$ positions along rows and columns of subimages and subfields. This was done in order to model the $x,y$ positions in terms of purely longitudinal and transverse image displacements. The advantage of this approach is simplicity as regards modeling but the disadvantage is that covariances can only be calculated from pairs of subimages and subfields that are on the same row or column of subapertures. A more appealing approach is to process all data without these restrictions. This can be done by calculating Fried's function $I(s/D_{\rm eff},\theta)$ for arbitrary angles $\theta$, defining the angle between a line connecting the two subaperture and that of the subfields measured. The {\em measured} $x,y-$ positions should first be rotated onto the line connecting the two apertures at the pupil plane. This angle rotates with height $h$. A subfield FOV defined by a separation $s$ at the pupil plane and field angles $\alpha$ and $\beta$ are projected at ($s+\alpha h, \beta h$) at a height $h$ and $\tan \theta = \beta h /(s+\alpha h)$ which at large heights approaches $\beta /\alpha$. Thus virtually every subfield and every subaperture needs a unique basis function for every height layer in the model. By pre-calculating $I(s/D_{\rm eff},\theta)$ and storing the result as a table, these basis functions can be calculated by interpolation.

\section{Conclusions}
The proposed method is based on measurements of {\em differential} measurements of seeing-induced image displacements, making it insensitive to telescope tracking errors, vibrations or residual errors from a tip-tilt mirror. The numerical computations of \citet{1975RaSc...10...71F} can be used to provide the theoretical covariance functions needed, requiring very small amounts of software development and avoiding the need for calculation of covariance functions via numerical turbulence simulations. The finite FOV used for wavefront sensing with solar granulation is accounted for in an approximate way by defining an effective subaperture diameter $D_{eff}$, increasing with height.

In terms of data collection, the proposed method is identical to the SLODAR method \citep{2002MNRAS.337..103W}, also employing Shack-Hartmann wavefront sensor data. However, the SLODAR method uses averages of measured image shifts from all subapertures for each of the (two) stars observed to eliminate the effects of telescope guiding errors. Eliminating the anisoplanatism introduced by this averaging requires a fairly elaborate analysis of the data \citep{2006MNRAS.369..835B}. In terms of analysis, the present method appears simpler.

Based on simulations and data processed from a single day of observations, we conclude that the proposed method combined with wavefront data over about 5$\times$5~arcsec subfields allows contributions to seeing from about 9-10 layers, stretching from the pupil up to 16-30~km distance from the telescope. At distances up to about 6~km, measurements with good S/N and a height resolution up to nearly 1~km appears possible. The 5.5~arcsec FOV used for wavefront sensing leads to poor sensitivity to high-altitude seeing and strongly reduced height resolution beyond 10~km. At a distance of 30~km, the FOV used corresponds to averaging wavefront information over a diameter of 80~cm. Our estimates of $r_0$ are very uncertain at this height and clearly the FOV needs to be reduced for seeing measurements at such large distances to be convincing. We also have established empirically that detecting seeing from high layers with the present system requires $r_0$ to be larger than approximately 7.5~cm for the ground layer and that the FOV should be such that the corresponding averaging area is smaller than $r_0$ at the high (10-30~km) layers. 

An important limitation is wavefront sensor noise, leading to bias in the measured covariances. By using image position measurements from 4 cross-correlation reference images, wavefront sensor noise is reduced and residual noise bias is estimated directly from the data and compensated for. However, the method used for estimation of noise bias relies on random errors from the 4 measurements to be independent and the method is furthermore ``blind'' to random errors that are the same for the 4 measurements. Noise propagation clearly needs further investigation.

The present method relies on wavefront tilts inferred from displacements of solar granulation images measured using cross--correlation techniques. The accuracy of these techniques for wavefront sensing is under investigation by means of simulations (L\"ofdahl, in preparation). A weakness of such conventional techniques is that the image displacement is assumed to be constant within the FOV used for cross-correlations. This corresponds to modeling the wavefront as having pure tip and tilt without high-order curvature. At the intersection of adjacent subfields, this corresponds to discontinuous gradients of the wavefront. Cross-correlations with overlapping sub-fields lead to multiple values of the wavefront gradient where subfields overlap. In addition to leading to inconsistencies and poor estimates of wavefront gradients, the use of conventional cross-correlation techniques should also lead to noisier measurements when differential seeing within the FOV is strong. A more satisfactory approach may be to use a 2D Fourier expansion of the image distortions, $\delta x= \delta x(\alpha,\beta)$ and $\delta y= \delta y(\alpha,\beta)$ over the entire FOV and to use that representation to calculate the {\em local} gradients of the wavefront. The highest order Fourier components would be limited by sampling and the FOV. This would be quite similar to fitting SH wavefront data to low-order Zernike or Karhunen-Loeve expansions, but over a rectangular FOV instead of over a round pupil. A simpler approach may be to use the 16 pixel FOV cross-correlations as initial estimates and then refine the estimate with a 8--12 pixel FOV, restricting the corrections of the image position
shifts to be small.

We believe that seeing measurements from the ground layer may be possible with the proposed method and appropriate noise bias compensation even when $r_0$ is somewhat less than 7.5~cm but that little or no information can be obtained from the higher layers in such conditions. The major problem is the relatively poor daytime seeing, limiting the quality and number of seeing estimates for the highest layers. Possibly, the Moon can be used as widefield wavefront sensor target for verification of day-time estimates of high-altitude seeing at night, but with obvious limitations in image scale, exposure time and S/N requiring careful consideration.  

\begin{acknowledgements}
The wavefront sensor optics was designed by Bo Lindberg at Lenstech AB and the wavefront sensor mechanics was designed and built by Felix Bettonvil and other members of the Dutch Open Telescope (DOT) team. The CCD camera software was written by Michiel van Noort. We are grateful for their help and assistance. We are also grateful for several valuable comments and suggestions by A. Tokovinin and M. Collados.

This research project has been supported by a Marie Curie Early Stage
Research Training Fellowship of the European Community's Sixth Framework
Programme under contract number MEST-CT-2005-020395: The USO-SP
International School for Solar Physics.

This work has been partially supported by the European Commission
through the collaborative project 212482 'EST: the large aperture European
Solar Telescope´ Design Study (FP7 - Research Infrastructures).

This work has been supported by a planning grant from the Swedish
Research Council.
\end{acknowledgements}

\bibliographystyle{aa}
\end{document}